\documentclass[journal,twoside,web]{ieeecolor}
\usepackage{tmp_st}
\usepackage{cite}
\usepackage{amsmath,amssymb,amsfonts}
\usepackage{algorithmic}
\usepackage{graphicx}
\usepackage{textcomp}
\usepackage{verbatim}
\usepackage{booktabs}

\def\BibTeX{{\rm B\kern-.05em{\sc i\kern-.025em b}\kern-.08em
    T\kern-.1667em\lower.7ex\hbox{E}\kern-.125emX}}
\newcommand{\TtwoCT}{T\textsubscript{2}$\rightarrow$CT}
\newcommand{\ToneCT}{T\textsubscript{1}$\rightarrow$CT}
\newcommand{\Tone}{T\textsubscript{1}}
\newcommand{\Ttwo}{T\textsubscript{2}}

\newcommand{\ToneTtwo}{T\textsubscript{1}$\rightarrow$T\textsubscript{2}}
\newcommand{\TonePD}{T\textsubscript{1}$\rightarrow$PD}
\newcommand{\TtwoTone}{T\textsubscript{2}$\rightarrow$T\textsubscript{1}}
\newcommand{\TtwoPD}{T\textsubscript{2}$\rightarrow$PD}
\newcommand{\PDTone}{PD$\rightarrow$T\textsubscript{1}}
\newcommand{\PDTtwo}{PD$\rightarrow$T\textsubscript{2}}

\newcommand{\TtwoFlair}{T\textsubscript{2}$\rightarrow$FLAIR}
\newcommand{\FlairTtwo}{FLAIR$\rightarrow$T\textsubscript{2}}

\newcommand{\ToneFlair}{T\textsubscript{1}$\rightarrow$FLAIR}
\newcommand{\FlairTone}{FLAIR$\rightarrow$T\textsubscript{1}}

\usepackage{tikz}
\usetikzlibrary{matrix,calc}
\usepackage{framed,multirow}
\usepackage{fixltx2e}
\usepackage{latexsym}
\usepackage{url}
\usepackage{xcolor}
\usepackage{lipsum}
\usepackage{hyperref}
\hypersetup{
    colorlinks=true,
    linkcolor=blue,
    filecolor=magenta,      
    urlcolor=cyan,
    pdftitle={SynDiff},
    }
\usepackage{gensymb}
\usepackage{caption}
\usepackage{comment}

\definecolor{newcolor}{rgb}{.8,.349,.1}
\def\baselinestretch{1.0}
\renewcommand{\arraystretch}{1.4}
\usepackage[math]{cellspace}
\def\SB#1{\textsubscript{#1}}

\definecolor{brightcerulean}{rgb}{0.11, 0.62, 0.74}
 \newcommand*{\revhl}{\textcolor{black}}

\makeatletter
\IEEEtriggercmd{\reset@font\normalfont\fontsize{7.9pt}{8.40pt}\selectfont}
\makeatother
\IEEEtriggeratref{1}

\begin{document}
\title{Unsupervised Medical Image Translation \\with Adversarial Diffusion Models}
\author{Muzaffer \"Ozbey$^*$ , Onat Dalmaz$^*$ , Salman UH Dar, Hasan A Bedel, \c{S}aban \"Ozturk, Alper G\"ung\"or,\\ and Tolga \c{C}ukur
\vspace{-1.25cm}
\\
\thanks{$^*$ M. Ozbey and O. Dalmaz contributed equally to this study.
This study was supported in part by TUBITAK BIDEB scholarships awarded to A. Bedel, O. Dalmaz, A. Gungor, and by TUBA GEBIP 2015 and BAGEP 2017 fellowships awarded to T. \c{C}ukur  (Corresponding author: Tolga \c{C}ukur).  M. Ozbey, O. Dalmaz, S.UH. Dar, H.A. Bedel, S. Ozturk, A. Gungor and T. \c{C}ukur are with the Department of Electrical and Electronics Engineering, and the National Magnetic Resonance Research Center (UMRAM), Bilkent University, Ankara, Turkey (e-mails: \{muzaffer, onat, salman, abedel, sozturk, agungor, cukur\}@ee.bilkent.edu.tr). S. Ozturk is also with Amasya University, Turkey. A. Gungor is also with ASELSAN Research Center, Turkey. The authors thank Mahmut Yurt for discussions on diffusion models in early phases of the study.}
}

\maketitle
\begin{abstract}
Imputation of missing images via source-to-target modality translation can improve diversity in medical imaging protocols. A pervasive approach for synthesizing target images involves one-shot mapping through generative adversarial networks (GAN). Yet, GAN models that implicitly characterize the image distribution can suffer from limited sample fidelity. Here, we propose a novel method based on adversarial diffusion modeling, SynDiff, for improved performance in medical image translation. To capture a direct correlate of the image distribution, SynDiff leverages a conditional diffusion process that progressively maps noise and source images onto the target image. For fast and accurate image sampling during inference, large diffusion steps are taken with adversarial projections in the reverse diffusion direction. To enable training on unpaired datasets, a cycle-consistent architecture is devised with coupled diffusive and non-diffusive modules that bilaterally translate between two modalities. Extensive assessments are reported on the utility of SynDiff against competing GAN and diffusion models in multi-contrast MRI and MRI-CT translation. Our demonstrations indicate that SynDiff offers quantitatively and qualitatively superior performance against competing baselines. 

\end{abstract}

\begin{IEEEkeywords}
medical image translation, synthesis, unsupervised, unpaired, adversarial, diffusion, generative  \vspace{-0.25cm}
\end{IEEEkeywords}

\bstctlcite{IEEEexample:BSTcontrol}
\section{Introduction}
Multi-modal imaging is key for comprehensive assessment of anatomy and function in the human body \cite{iglesias2013}. Complementary tissue information captured by individual modalities serve to improve diagnostic accuracy and performance in downstream imaging tasks \cite{lee2017}. Unfortunately, broad adoption of multi-modal protocols is fraud with challenges due to economic and labor costs \cite{ye2013,huynh2015,jog2017,joyce2017}. Medical image translation is a powerful solution that involves synthesis of a missing target modality under guidance from an acquired source modality \cite{cordier2016,wu2016,zhao2017,huang2018}. This recovery is an ill-conditioned problem given nonlinear variations in tissue signals across modalities \cite{roy2013,alexander2014,huang2017}. At this juncture, learning-based methods are offering performance leaps by incorporating nonlinear data-driven priors to improve problem conditioning \cite{hien2015,vemulapalli2015,sevetlidis2016,nie2016}.

Learning-based image translation involves network models trained to capture a prior on the conditional distribution of target given source images \cite{bowles2016,chartsias2018,wei2019}. In recent years, generative adversarial network (GAN) models have been broadly adopted for translation tasks, given their exceptional realism in image synthesis \cite{dar2019image,yu2018,nie2018,armanious2019,lee2019,li2019}. A discriminator that captures information regarding the target distribution concurrently guides a generator to perform one-shot mapping from the source onto the target image \cite{yu2019,mmgan,wang2020,zhou2020,lan2020}. Based on this adversarial mechanism, state-of-the-art results have been reported with GANs in numerous translation tasks including synthesis across MR scanners \cite{nie2018}, multi-contrast MR synthesis \cite{dar2019image,lee2019,yu2019,yurt2021mustgan}, and cross-modal synthesis \cite{yang2018,jin2018,resvit}.    

While powerful, GAN models indirectly characterize the distribution of the target modality through a generator-discriminator interplay without evaluating likelihood \cite{pix2pix}. Such implicit characterization is potentially amenable to learning biases, including premature convergence and mode collapse. \revhl{Moreover, GAN models commonly employ a rapid one-shot sampling process without intermediate steps, inherently limiting the reliability of the mapping performed by the network. In turn, these issues can limit the quality and diversity of synthesized images \cite{DiffBeatsGAN}. As a promising alternative, recent computer vision studies have adopted diffusion models based on explicit likelihood characterization and a gradual sampling process} to improve sample fidelity in unconditional generative modeling tasks \cite{DDPM,DiffBeatsGAN}. However, the potential of diffusion methods in medical image translation remains largely unexplored, partly owing to the computational burden of image sampling and difficulties in unpaired training of regular diffusion models \cite{DDPM}. 

\begin{figure*}[t]
\begin{minipage}{0.325\textwidth}
\caption{\textbf{a)} Regular diffusion models gradually transform between actual image samples for the target modality ($\boldsymbol{x}_0$) and isotropic Gaussian noise ($\boldsymbol{x}_T$) in $T$ steps, with $T$ on the order of thousands. \revhl{Each forward step (right arrows)} adds noise to the current sample to create a noisier sample with forward transition probability $q\left( \boldsymbol{x}_{t+1}|\boldsymbol{x}_{t} \right)$. \revhl{Each reverse step (left arrows)} suppresses the added noise to create a denoised sample. For image translation, a source modality ($\boldsymbol{y}$) can also be provided as conditioning input to the reverse steps resulting in a reverse transition probability $q\left( \boldsymbol{x}_{t-1}|\boldsymbol{x}_{t},\boldsymbol{y} \right)$ \revhl{assumed to be Gaussian, and operationalized via a neural network $p_{\boldsymbol{\theta}}\left( \boldsymbol{x}_{t-1}|\boldsymbol{x}_{t},\boldsymbol{y} \right)$ that estimates its mean.} \textbf{b)} The proposed adversarial diffusion model performs fast transformation between $\boldsymbol{x}_0$ and $\boldsymbol{x}_T$ in $T/k$ steps, with step size $k \gg 1$. \revhl{Each forward step adds a greater amount to noise to compensate for large $k$, breaking apart the normality assumption for reverse transition probabilities $q\left( \boldsymbol{x}_{t-k}|\boldsymbol{x}_{t},\boldsymbol{y} \right)$. To improve accuracy, reverse diffusion steps are operationalized via a novel adversarial projector that uses a generator $G_{\boldsymbol{\theta}}$ and a discriminator $D_{\boldsymbol{\theta}}$. $G_{\boldsymbol{\theta}}$ first produces an estimate of the target image $\Tilde{\boldsymbol{x}}_0$ given $\boldsymbol{x}_t$ and $\boldsymbol{y}$, and a denoised image sample $\hat{\boldsymbol{x}}_{t-k}$ is then synthesized from the denoising distribution $q\left( \boldsymbol{x}_{t-k}|\boldsymbol{x}_{t},\Tilde{\boldsymbol{x}}_0 \right)$.} Meanwhile, $D_{\boldsymbol{\theta}}$ distinguishes between actual ($\boldsymbol{x}_{t-k}$) and synthetic samples ($\hat{\boldsymbol{x}}_{t-k}$) for the denoised image.}
\label{fig:advdiff}
\end{minipage}
\begin{minipage}{0.675\textwidth}
\centerline{\includegraphics[width=0.92\columnwidth]{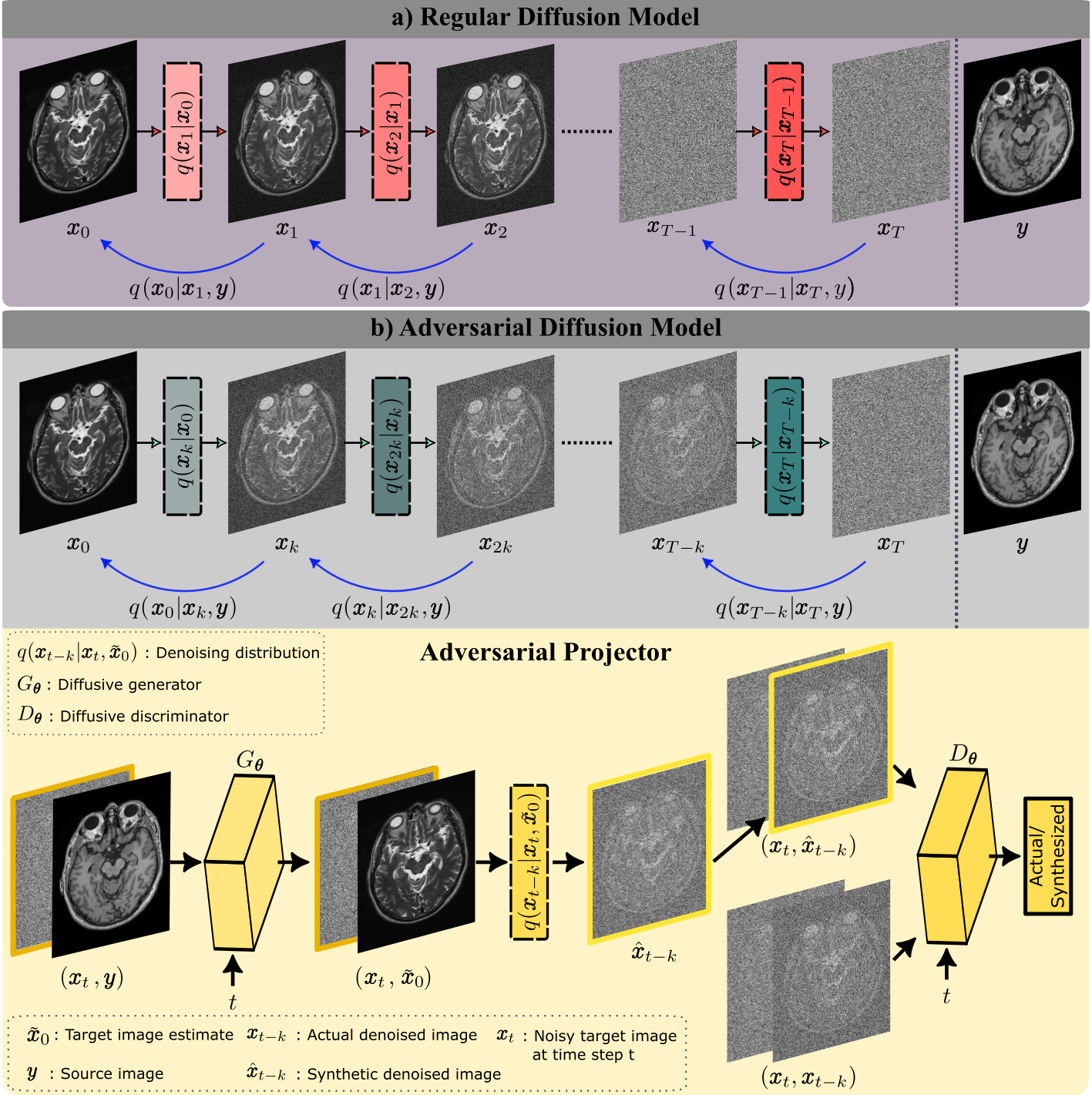}}
\end{minipage}
\hfill
\end{figure*}

Here, we propose a novel adversarial diffusion model for medical image synthesis, SynDiff, to perform efficient and high-fidelity modality translation (Fig. \ref{fig:advdiff}). Given the source image, SynDiff leverages conditional diffusion to generate the target image. Unlike regular diffusion models, SynDiff adopts a fast diffusion process with large step size for efficiency. Accurate sampling in reverse diffusion steps is achieved by a novel \revhl{source-conditional adversarial projector that denoises the target image sample with guidance from the source image}. To enable unsupervised learning, a cycle-consistent architecture is devised with bilaterally coupled diffusive and non-diffusive processes between the two modalities (Fig. \ref{fig:syndiff}). Comprehensive demonstrations are performed for multi-contrast MRI and MRI-CT translation. Our results clearly indicate the superiority of SynDiff against competing GAN and diffusion models. Code for SynDiff is publicly available at \href{https://github.com/icon-lab/SynDiff}{https://github.com/icon-lab/SynDiff}. 

\vspace{2mm}
\textcolor{cyan}{\textbf{Contributions}}
\begin{itemize}
    \item We introduce the first adversarial diffusion model in the literature for high-fidelity medical image synthesis.  
    \item We introduce the first \revhl{diffusion-based method for unsupervised medical image translation} that enables training on unpaired datasets of source-target modalities. 
    \item We propose a novel \revhl{source-conditional} adversarial projector to capture reverse transition probabilities over large step sizes for efficient image sampling. 
\end{itemize}

\section{Related Work}
To translate medical images, conditional GANs perform one-shot source-to-target mapping via a generator trained using an adversarial loss \cite{nie2018}. Adversarial loss terms are known to improve sensitivity to high-frequency details in tissue structure over canonical pixel-wise losses \cite{dar2019image}. As such, GAN-based translation has been broadly adopted in many applications. Augmenting adversarial with pixel-wise losses, a first group of studies considered supervised training on paired sets of source-target images matched across subjects \cite{armanious2019,mmgan,li2019,yu2019,zhou2020,wang2020}. For improved flexibility, other studies proposed cycle-consistency or mutual information losses to enable unsupervised learning from unpaired data \cite{chartsias2017,wolterink2017,hiasa2018,yang2018,dar2019image,sohail2019,ge2019,dong2019}. In general, enhanced spatial acuity and realism have been reported in target images synthesized with GANs when compared to vanilla convolutional models \cite{dar2019image}. That said, several problems can arise in GAN models, including \revhl{lower mapping reliability for the one-shot sampling process \cite{DiffBeatsGAN}}, premature convergence of the discriminator before the generator is properly trained \cite{lan2020}, and poor representational diversity due to mode collapse \cite{pix2pix}. In turn, these problems can lower sample quality and diversity, limiting generalization performance of GAN-based image translation. 

As a recent alternative to GANs, deep diffusion models have received interest for generative modeling tasks in computer vision \cite{DDPM,DiffBeatsGAN}. Starting from a pure noise sample, diffusion models generate image samples from a desired distribution through repetitive denoising. Denoising is performed via a neural network architecture trained to maximize a correlate on data likelihood. Due to the gradual stochastic sampling process and explicit likelihood characterization, diffusion models can \revhl{improve the reliability of the network mapping to} offer enhanced sample quality and diversity. Given this potential, diffusion-based methods have recently been adopted for unimodal imaging tasks such as image reconstruction \cite{jalaln2021nips,chung2022media,song2022solving,adadiff,denoisingmri}, unconditional image generation \cite{CardosoLatentDiffusion}, and anomaly detection \cite{diffusion_anomaly_detection,cardoso_anomaly_diffusion}. Nevertheless, these methods are typically based on unconditional diffusion processes devised to process single-modality images. Furthermore, current methods often involve vanilla diffusion models that rely on a large number of inference steps for accurate image generation. This prolonged sampling process introduces computational challenges in adoption of diffusion models.

Here, we propose a novel adversarial diffusion model for improved efficiency and performance in medical image translation. Unlike recent imaging methods based on unconditional diffusion, SynDiff leverages \revhl{conditional diffusion where a source-contrast image anatomically guides the reverse diffusion process. A novel source-conditional adversarial projector is employed for efficient and accurate image sampling over few large diffusion steps.} Furthermore, a novel cycle-consistent architecture is introduced combining diffusive and non-diffusive modules to enable unsupervised training. To our knowledge, SynDiff is the first adversarial diffusion model for medical image synthesis, and the first \revhl{diffusion-based method for unsupervised medical image translation} in the literature. Based on these unique advances, we provide \revhl{the first demonstrations of unsupervised translation in multi-contrast MRI and multi-modal MRI-CT based on diffusion modeling}.  

\revhl{Few recent studies have considered improvements on vanilla diffusion models with partially related aims to our proposed method. A study on natural image generation has used an adversarial diffusion model, DDGAN, to improve efficiency in reverse diffusion steps \cite{DiffNvidia}. DDGAN is based on an unconditional diffusion process that generates random images starting from noise; and it uses an adversarial generator for reverse diffusion without guidance from a source image. In contrast, SynDiff is based on a conditional diffusion process that translates between source- and target-images of an anatomy. It uses a source-conditional adversarial projector for reverse diffusion to synthesize target images with anatomical correspondence to a guiding source image. Besides the diffusive module, SynDiff also embodies a non-diffusive module to permit unsupervised image translation. A study on unsupervised translation of natural images has proposed a non-adversarial diffusion model, UNIT-DDPM \cite{sasaki2021unit}. Based on the notion that source-target modalities share a latent space, UNIT-DDPM uses parallel diffusion processes to simultaneously generate samples for both modalities in a large number of reverse steps; and the noisy source-image samples drawn from the source diffusion process are used to condition the generation of target images in the target diffusion process. In contrast, SynDiff uses an adversarial projector for efficient sampling in few steps; and it leverages source-image estimates that are produced by a non-diffusive module to provide high-quality anatomical guidance for synthesis of target images.} A recent study has independently considered a conditional score-based method, UMM-CGSM, for imputation of missing sequences in a multi-contrast MRI protocol \cite{meng_arxiv_2022}. UMM-CGSM uses a non-adversarial model with relatively large number of inference steps; and it performs supervised training on paired datasets of source-target images. In contrast, SynDiff adopts an adversarial diffusion model for efficient sampling over few steps; and it can perform unsupervised learning.

\begin{table}[ht!]
\caption{\revhl{Description of variables related to images, diffusion processes, networks and probability distributions. Throughout the manuscript, vectorial quantities are annotated in bold font.}}
\centering
\resizebox{0.975\columnwidth}{!}{%
\begin{tabular}{ll}
 \hline
 \textbf{Images} & {}  \\
  \hline
  $\boldsymbol{x}_0$   & Actual target-image sample  \\
  \hline
  $\boldsymbol{x}_t$   & Noisy target-image sample at time step $t$  \\
  \hline
  \multirow{2}{*}{$\boldsymbol{x}_T$}   & Noisy target-image sample at time step $T$, \\
                          & (i.e., drawn from isotropic Gaussian distribution)  \\
  \hline
  $\boldsymbol{y}$   & Guiding source image  \\
  \hline
  $\boldsymbol{x}_{t-k}$ & Actual target-image sample at time step $t-k$  \\
 \hline
  $\hat{\boldsymbol{x}}_{t-k}$   & Synthesized target-image sample at time step $t-k$  \\
  \hline
  $\boldsymbol{x}_0^A$, $\boldsymbol{x}_0^B$   & Unpaired training images from modalities A and B  \\
  \hline
  $\tilde{\boldsymbol{y}}^A, \tilde{\boldsymbol{y}}^B$   & Source images estimated by non-diffusive module  \\
  \hline
  $\breve{\boldsymbol{x}}_0^{A},\breve{\boldsymbol{x}}_0^{B}$   & Target images synthesized by non-diffusive module\\
  \hline
  $\hat{\boldsymbol{x}}_0^{A},\hat{\boldsymbol{x}}_0^{B}$   & Target images synthesized by diffusive module\\
  \hline
  \hline
  \textbf{Regular diffusion} & {} \\ 
  \hline
    $\beta_t$   & Noise variance for regular diffusion at time step $t$  \\
  \hline
   $\boldsymbol{\epsilon}$   & Standard normal random vector  \\
  \hline
  $\boldsymbol{\mu}(\boldsymbol{x}_t, t), \mathbf{\boldsymbol{\Sigma}}(\boldsymbol{x}_t, t)$   & Network estimates for mean and covariance \\ &  of the conditional distribution of $\boldsymbol{x}_{t-1}$ given $\boldsymbol{x}_{t}$   \\
  \hline
  $\psi_{t}$   & $1-\beta_{t}$  \\
  \hline
  $\overline{\psi}_{t}$   & $\prod_{r=[0,1,..,t]}\psi_{r}$  \\
  \hline
  $\boldsymbol{\epsilon}_{\theta}(\boldsymbol{x}_t, t)$   & Network estimate for the added noise at time step $t$ \\
  \hline
  \hline
  \textbf{Adversarial diffusion} & {} \\ 
  \hline
  $k$   & Step size for fast diffusion \\
  \hline
  $\gamma_t$   & Noise variance for fast diffusion at time step $t$  \\
  \hline
  $\overline{\beta}_{\text{min}}, \overline{\beta}_{\text{max}}$   & Parameters that control the progression of noise variance  \\
  \hline
$\boldsymbol{f}_i$   & Feature maps in the $i$th subblock of the diffusive generator  \\
  \hline
  $\boldsymbol{m}$   & Learnable temporal embedding added onto feature maps \\ & to encode the time step $t$  \\
   \hline
  $\alpha_{t}$   & $1-\gamma_{t}$  \\
  \hline
  $\overline{\alpha}_{t}$   & $\prod_{r=[0,k,..,t]}\alpha_{r}$  \\
  \hline
    \hline
  \textbf{Networks} & {} \\ 
  \hline
  $G^A_{\boldsymbol{\phi}},D^A_{\boldsymbol{\phi}}$   & Non-diffusive generator-discriminator pair for learning \\ & to estimate a source image $\tilde{\boldsymbol{y}}^A$ given $\boldsymbol{x}_0^B$   \\
  \hline
  $G^B_{\boldsymbol{\phi}},D^B_{\boldsymbol{\phi}}$   & Non-diffusive generator-discriminator pair for learning \\ & to estimate a source image $\tilde{\boldsymbol{y}}^B$ given $\boldsymbol{x}_0^A$   \\
  \hline
  $G^A_{\boldsymbol{\theta}},D^A_{\boldsymbol{\theta}}$   & Diffusive generator-discriminator pair for learning \\ & to synthesize a target image $\hat{\boldsymbol{x}}^A_{t-k}$ given $\boldsymbol{x}^A_{t}$   \\
  \hline
  $G^B_{\boldsymbol{\theta}},D^B_{\boldsymbol{\theta}}$   & Diffusive generator-discriminator pair for learning \\ & to synthesize a target image $\hat{\boldsymbol{x}}^B_{t-k}$ given $\boldsymbol{x}^B_{t}$   \\
  \hline
  \hline
  \multicolumn{2}{l}{\textbf{Probability distributions}} \\
  \hline
  $q(\boldsymbol{x}_0)$   & Actual image distribution  \\
  \hline
  $q(\boldsymbol{x}_t|\boldsymbol{x}_{t-1})$   & Forward transition probability  \\
  \hline
  $q(\boldsymbol{x}_{t-1}|\boldsymbol{x}_t)$   & Reverse transition probability  \\
  \hline
  $p_\theta(\boldsymbol{x}_{t-1}|\boldsymbol{x}_t)$   & Network estimate for reverse
  transition probability\\
  \hline
  $q(\boldsymbol{x}_{t-k}|\boldsymbol{x}_t,\boldsymbol{y})$   & Reverse transition probability for fast conditional \\ &diffusion with step size $k \gg 1$ \\
  \hline
  $p_{\boldsymbol{\theta}}(\boldsymbol{x}_{t-k}|\boldsymbol{x}_t,\boldsymbol{y})$   & Network estimate for reverse transition probability \\& in fast conditional diffusion with step size $k \gg 1$ \\
  \hline
\end{tabular}
}
\end{table}

\section{Theory}
\subsection{Denoising Diffusion Models}
Regular diffusion models map between pure noise samples and actual images through a gradual process over $T$ time steps (Fig. \ref{fig:advdiff}a). In the forward direction, a small amount of Gaussian noise is added repeatedly onto an input image $\boldsymbol{x}_0 \sim q(\boldsymbol{x}_0)$ to obtain a sample $\boldsymbol{x}_T$ from an isotropic Gaussian distribution for sufficiently large $T$. Forward diffusion forms a Markov chain where the mapping from $\boldsymbol{x}_{t-1}$ to $\boldsymbol{x}_t$ and the respective forward transition probability are:
\begin{eqnarray}
&& \boldsymbol{x}_{t}=\sqrt{1-\beta_{t}}\boldsymbol{x}_{t-1}+\sqrt{\beta_{t}}\boldsymbol{\epsilon},\quad{}\boldsymbol{\epsilon}\sim \mathcal{N}\left( \boldsymbol{0},\boldsymbol{I} \right) \\
&& q\left( \boldsymbol{x}_{t}|\boldsymbol{x}_{t-1} \right)=\mathcal{N}\left( \boldsymbol{x}_{t}; \sqrt{1-\beta_{t}}\boldsymbol{x}_{t-1},\beta_{t}\boldsymbol{I} \right)
\end{eqnarray}
where $\beta_t$ is noise variance, $\boldsymbol{\epsilon}$ is added noise, $\mathcal{N}$ is a Gaussian distribution, $\boldsymbol{I}$ is an identity covariance matrix. Reverse diffusion also forms a Markov chain from $\boldsymbol{x}_T$ onto $\boldsymbol{x}_0$, albeit each step aims to gradually denoise the samples. Under large $T$ and small $\beta_t$, the reverse transition probability between $\boldsymbol{x}_{t-1}$ and $\boldsymbol{x}_t$ can be approximated as a Gaussian distribution \cite{sohl2015deep, feller1949theory}: 
\begin{equation} \label{eq:backward}
	q(\boldsymbol{x}_{t-1}|\boldsymbol{x}_{t}) := \mathcal{N}(\boldsymbol{x}_{t-1}; \boldsymbol{\mu}(\boldsymbol{x}_t, t), \mathbf{\boldsymbol{\Sigma}}(\boldsymbol{x}_t, t))
\end{equation}

Diffusion models typically operationalize each reverse diffusion step as mapping through a neural network that provides estimates for $\boldsymbol{\mu}$ and/or $\boldsymbol{\Sigma}$. Training is then performed by minimizing a variational bound on log-likelihood: 
\begin{equation} \label{eq:lvlb}
    L_{vb} = \mathbb{E}_{q(\boldsymbol{x}_{0:T})} \left[ \, log \, \frac{p_{\theta}(\boldsymbol{x}_{0:T})}{q(\boldsymbol{x}_{1:T} | \boldsymbol{x}_0)} \right] \le \mathbb{E}_{q(\boldsymbol{x}_{0})} \left[ \, log \, p_{\boldsymbol{\theta}}(\boldsymbol{x}_0) \right] 
\end{equation}
where $\mathbb{E}_q$ denotes expectation over $q$, $p_{\boldsymbol{\theta}}$ is the network parametrization of \revhl{the joint distribution of input variables, $\boldsymbol{\theta}$ are network parameters, $\boldsymbol{x}_{0:T}$ denote the collection of image samples between time steps $0$ and $T$, and $\boldsymbol{x}_{1:T} | \boldsymbol{x}_0$ denote image samples between time steps $1$ and $T$ conditioned on the sample at time step $0$}. The bound can be decomposed as:
\begin{eqnarray}
\label{eq:Lall}
\hspace{-3.2mm}
L_{vb} &=& log \, p_{\boldsymbol{\theta}} (\boldsymbol{x}_0 | \boldsymbol{x}_1) \nonumber \\ 
&&- \sum_{t=1}^{T} KL(q(\boldsymbol{x}_{t-1} | \boldsymbol{x}_t , \boldsymbol{x}_0) \, || \, p_{\boldsymbol{\theta}}(\boldsymbol{x}_{t-1}|\boldsymbol{x}_t)) 
\end{eqnarray}
where $KL$ denotes Kullback-Leibler divergence, and $KL(q(\boldsymbol{x}_T | \boldsymbol{x}_0) \, || \, p(\boldsymbol{x}_T))$ is omitted as it does not depend on $\boldsymbol{\theta}$. A common parametrization omits $\boldsymbol{\Sigma}$ to focus on $\boldsymbol{\mu}$:
\begin{equation} \label{eq:mean}
    \boldsymbol{\mu}_{\boldsymbol{\theta}}(\boldsymbol{x}_t, t) = \frac{1}{\sqrt{\psi_t}} \left( \boldsymbol{x}_t - \frac{\beta_t}{\sqrt{1 - \overline{\psi}_t}} \boldsymbol{\epsilon}_{\theta}(\boldsymbol{x}_t, t) \right)
\end{equation}
\revhl{where \( \psi_{t}=1-\beta_{t}\) and \( \overline{\psi}_{t}=\prod_{r=[0,1,..,t]}\psi_{r}\)}. In Eq. \ref{eq:mean}, $\boldsymbol{\mu}_{\boldsymbol{\theta}}$ can be derived if the network is used to estimate the added noise $\boldsymbol{\epsilon}$ by minimizing the following loss \cite{ho2020denoising}:
\begin{equation} \label{eq:Lerr}
    L_{err} = \mathbb{E}_{t, \boldsymbol{x}_0, \boldsymbol{\epsilon}} \left[  \| \boldsymbol{\epsilon} - \boldsymbol{\epsilon}_{\boldsymbol{\theta}}(\sqrt{\overline{\alpha}}_t \boldsymbol{x}_0 + \sqrt{1-\overline{\alpha}}_t \boldsymbol{\epsilon}, t) \|_2^2 \right]
\end{equation}
where $t$, $\boldsymbol{x}_0$ and $\boldsymbol{\epsilon}$ are sampled from \revhl{the discrete uniform distribution} $\mathcal{U}(0,T)$, $q(\boldsymbol{x}_0)$ and $\mathcal{N}\left( \boldsymbol{0},\mathrm{\boldsymbol{I}} \right)$, respectively. During inference, reverse diffusion steps are performed starting from a random sample $\boldsymbol{x}_T \sim \mathcal{N}(\boldsymbol{0}, \boldsymbol{I})$. For each step $t \in T...1$, $\boldsymbol{\mu}$ is derived using Eq. \ref{eq:mean} based on the network estimate $\boldsymbol{\epsilon}_{\boldsymbol{\theta}}$, and $\boldsymbol{x}_{t-1}$ is sampled based on Eq. \ref{eq:backward}.

\begin{figure*}[t]
\begin{minipage}{0.7\textwidth}
\centerline{\includegraphics[width=0.94\columnwidth]{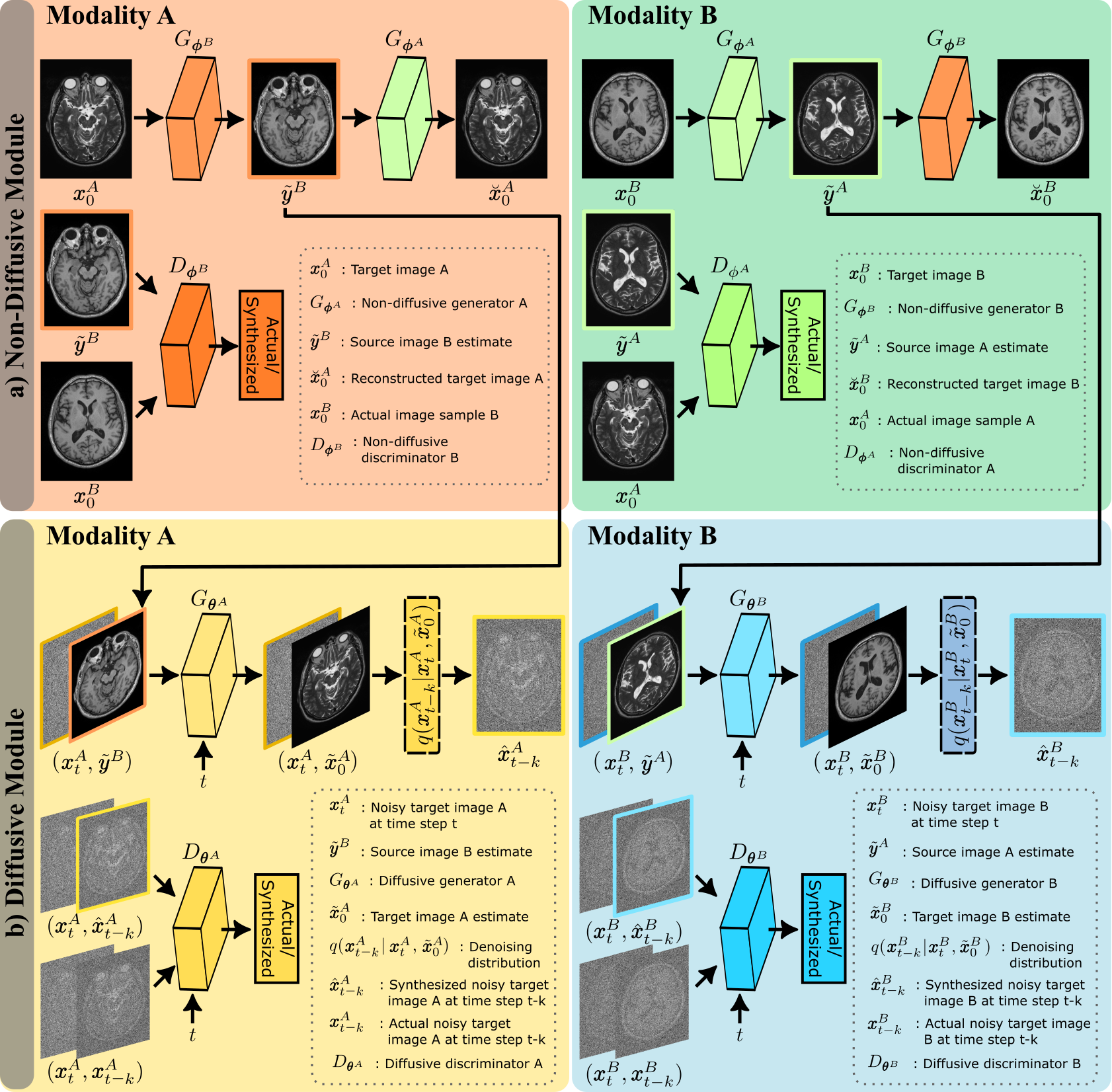}}
\end{minipage}
\begin{minipage}{0.3\textwidth}
\caption{For unsupervised learning, SynDiff leverages a cycle-consistent architecture that bilaterally translates between two modalities ($A$, $B$). For synthesizing a target image $\hat{\boldsymbol{x}}_0^A$ of modality $A$, the diffusive module in Fig. \ref{fig:advdiff}b requires guidance from a source image $\boldsymbol{y}^B$ of modality $B$ for the same anatomy. However, a paired source image of the same anatomy might be unavailable in the training set. To enable training on unpaired images, SynDiff uses a non-diffusive module to first estimate a paired source image $\tilde{\boldsymbol{y}}^B$ from $\boldsymbol{x}_0^A$. \revhl{Similarly, for synthesizing a target image $\hat{\boldsymbol{x}}_0^B$ of modality $B$ with the diffusive module, the non-diffusive module first estimates a paired source image $\tilde{\boldsymbol{y}}^A$ from $\boldsymbol{x}_0^B$.} \textbf{a)} To do this, the non-diffusive module comprises two generator-discriminator pairs ($G_{{\boldsymbol{\phi}}^{A,B}}$, $D_{{\boldsymbol{\phi}}^{A,B}}$) that generate initial translation estimates for $\boldsymbol{x}_0^A \rightarrow \tilde{\boldsymbol{y}}^B$ (orange) and $\boldsymbol{x}_0^B \rightarrow \tilde{\boldsymbol{y}}^A$ (green). \textbf{b)} These initial translation estimates $\tilde{\boldsymbol{y}}^{A,B}$ are then used as guiding source-modality images in the diffusive module. For cycle-consistent learning, the diffusive module also comprises two generator-discriminator pairs ($G_{{\boldsymbol{\theta}}^{A,B}}$, $D_{{\boldsymbol{\theta}}^{A,B}}$) to generate denoised image estimates for $(\boldsymbol{x}_t^A,\tilde{\boldsymbol{y}}^B,t) \rightarrow \hat{\boldsymbol{x}}_{t-k}^A$ (yellow) and $(\boldsymbol{x}_t^B,\tilde{\boldsymbol{y}}^A,t) \rightarrow \hat{\boldsymbol{x}}_{t-k}^B$ (blue).}
\label{fig:syndiff}
\end{minipage}
\hfill
\end{figure*}

\subsection{SynDiff}
Here, we introduce a novel diffusion model for efficient, high-fidelity translation between source and target modalities of a given anatomy. \revhl{SynDiff uses a diffusive module equipped with a source-conditional adversarial projector} for fast and accurate reverse diffusion sampling (Fig. \ref{fig:advdiff}b). \revhl{It also employs a non-diffusive module for estimating source images paired with corresponding target images, so as to enable unsupervised learning (Fig. \ref{fig:syndiff}). The adversarial diffusion process that forms the basis of the diffusive module, the network architecture, and the learning procedures for SynDiff are detailed below.}

\subsubsection{Adversarial Diffusion Process}
\label{sec:adp}
Regular diffusion models prescribe relatively large $T$ such that the step size is sufficiently small to satisfy the normality assumption in Eq. \ref{eq:backward}, but this limits efficiency in image generation. Here, we instead propose fast diffusion with the following forward steps:
\begin{eqnarray}
\label{eq:fastforward}
    &&\boldsymbol{x}_t = \sqrt{1-\gamma_t}\boldsymbol{x}_{t-k} + \sqrt{\gamma_t} \boldsymbol{\epsilon} \\
    &&q(\boldsymbol{x}_t|\boldsymbol{x}_{t-k}) = \mathcal{N} \left( \boldsymbol{x}_t; \sqrt{1-\gamma_t}\boldsymbol{x}_{t-k}, \gamma_t \boldsymbol{I}  \right)
\end{eqnarray}
where $k \gg 1$ is step size. The noise variance $\gamma_t$ is set as:
\begin{equation}
    \gamma_t = 1 - \mathrm{e}^{\overline{\beta}_{\text{min}}\frac{k}{T} -(\overline{\beta}_{\text{max}}- \overline{\beta}_{\text{min}}) \frac{2tk-k^2}{2T^2}}
\end{equation}
$\overline{\beta}_{\text{min}}$ and $\overline{\beta}_{\text{max}}$ control the progression of noise variance in an exponential schedule \cite{song2020score}. 

Guidance from a source image ($\boldsymbol{y}$) is available during medical image translation, so a conditional process is proposed in the reverse diffusion direction. Note that, for $k \gg 1$, there is no closed form expression for $q(\boldsymbol{x}_{t-k} | \boldsymbol{x}_t, \boldsymbol{y})$ and the normality assumption used to compute Eq. \ref{eq:lvlb} breaks down \cite{DDPM}. 
Here we introduce a novel \revhl{source-conditional} adversarial projector to capture the complex transition probability $q(\boldsymbol{x}_{t-k}| \boldsymbol{x}_t, \boldsymbol{y})$ for large $k$ in our conditional diffusion model, as inspired by a recent report on unconditional generation of natural images \revhl{using adversarial learning to capture $q(\boldsymbol{x}_{t-k}| \boldsymbol{x}_t)$} \cite{DiffNvidia}. In SynDiff, a conditional generator $G_{\boldsymbol{\theta}}(\boldsymbol{x}_t,\boldsymbol{y},t)$ performs gradual denoising in each reverse step to synthesize $\hat{\boldsymbol{x}}_{t-k} \sim p_{\boldsymbol{\theta}}(\boldsymbol{x}_{t-k}|\boldsymbol{x}_t, \boldsymbol{y})$. \revhl{$G_{\boldsymbol{\theta}}$ receives the image pair ($\boldsymbol{x}_t$,$\boldsymbol{y}$) as a two-channel input, and it extracts intermediate feature maps $\boldsymbol{f}_i$ where $i \in [1,...,N]$ is the subblock index in an encoder-decoder structure \cite{song2020score}. A learnable temporal embedding $\boldsymbol{m}$ is computed given $t$, and this embedding is added as a channel-specific bias term onto the feature maps in each subblock \cite{song2020score}: $\boldsymbol{f}'_i = \boldsymbol{f}_i + \boldsymbol{m}$.} Meanwhile, a discriminator $D_{\boldsymbol{\theta}}(\{\hat{\boldsymbol{x}}_{t-k} \mbox{ or } \boldsymbol{x}_{t-k}\},\boldsymbol{x}_t,t)$ distinguishes samples drawn from estimated versus true denoising distributions ($p_{\boldsymbol{\theta}}(\boldsymbol{x}_{t-k}|\boldsymbol{x}_t, \boldsymbol{y})$ vs. $q(\boldsymbol{x}_{t-k}| \boldsymbol{x}_t, \boldsymbol{y})$). \revhl{$D_{\theta}$ receives either ($\boldsymbol{x}_t$,$\hat{\boldsymbol{x}}_{t-k}$) or ($\boldsymbol{x}_t$,${\boldsymbol{x}}_{t-k}$) as a two-channel input. The temporal embedding $\boldsymbol{m}$ is also added as a bias term onto the feature maps across $D_{\boldsymbol{\theta}}$.} A non-saturating adversarial loss is adopted for $G_{\boldsymbol{\theta}}$ \cite{goodfellow2014generative}: 
\begin{eqnarray} \label{eq:G-loss} 
   L_{G_{\boldsymbol{\theta}}} = \mathbb{E}_{t,q(\boldsymbol{x}_t|\boldsymbol{x}_0,\boldsymbol{y}),p_{\boldsymbol{\theta}}(\boldsymbol{x}_{t-k}|\boldsymbol{x}_t, \boldsymbol{y})} [-log(D_{\boldsymbol{\theta}}(\hat{\boldsymbol{x}}_{t-k}))]
\end{eqnarray}
where $t\sim \mathcal{U}(\{0,k, ..., T\})$, and the discriminator arguments are abbreviated for brevity. $D_{\boldsymbol{\theta}}$ also adopts a non-saturating adversarial loss with gradient penalty \cite{mescheder2018training}:
\begin{align} \label{eq:D-loss} 
&&L_{D_{\boldsymbol{\theta}}} = \mathbb{E}_{t,q(\boldsymbol{x}_t|\boldsymbol{x}_0, \boldsymbol{y})} \left [ \mathbb{E}_{q(\boldsymbol{x}_{t-k}|\boldsymbol{x}_t, \boldsymbol{y})} \left[ -log(D_{\boldsymbol{\theta}}(\boldsymbol{x}_{t-k})) \right] \right.  \notag \\
&&+ \mathbb{E}_{p_{\boldsymbol{\theta}}(\boldsymbol{x}_{t-k}|\boldsymbol{x}_t, \boldsymbol{y})} [ -log(1-D_{\boldsymbol{\theta}}(\hat{\boldsymbol{x}}_{t-k})) ]  \notag \\ 
&&\left. + \eta \mathbb{E}_{q(\boldsymbol{x}_{t-k}|\boldsymbol{x}_t,\boldsymbol{y})} \left\| \nabla_{\boldsymbol{x}_{t-k}} D_{\boldsymbol{\theta}}(\boldsymbol{x}_{t-k})  \right\|^2_2  \right]  
\end{align}
where $\eta$ is the weight for the gradient penalty.

Evaluation of Eqs. \ref{eq:G-loss}-\ref{eq:D-loss} require sampling from $q(\boldsymbol{x}_{t-k}|\boldsymbol{x}_t, \boldsymbol{y})$ that is unknown. \revhl{Yet, since $\boldsymbol{x}_0$ and $\boldsymbol{y}$ are non-linearly related images of the same anatomy and} $\boldsymbol{x}_t$ is conditionally independent of $\boldsymbol{y}$ given $\boldsymbol{x}_0$, \revhl{the reverse transition probability can be expressed as} $q(\boldsymbol{x}_{t-k}|\boldsymbol{x}_t, \boldsymbol{x}_0, \boldsymbol{y}) = q(\boldsymbol{x}_{t-k}|\boldsymbol{x}_t, \boldsymbol{x}_0)$ \cite{DDPM}. Bayes' rule can then be used to express the denoising distribution in terms of forward transition probabilities:
\begin{align}
\label{eq:qsample}
    q(\boldsymbol{x}_{t-k}|\boldsymbol{x}_t,\boldsymbol{x}_0) = q(\boldsymbol{x}_t|\boldsymbol{x}_{t-k}, \boldsymbol{x}_0)\frac{q(\boldsymbol{x}_{t-k}|\boldsymbol{x}_0)}{q(\boldsymbol{x}_{t}|\boldsymbol{x}_0)} 
\end{align}
Using Eq. \ref{eq:fastforward}, it can then be shown that $q(\boldsymbol{x}_{t-k}|\boldsymbol{x}_t,\boldsymbol{x}_0) = \mathcal{N}(\boldsymbol{x}_{t-k}; \overline{\boldsymbol{\mu}}(\boldsymbol{x}_t, \boldsymbol{x}_0), \overline{\gamma}\boldsymbol{I})$ with the following parameters: 
\begin{align}
    \overline{\boldsymbol{\mu}}=\frac{\sqrt{\overline{\alpha}_{t-k}}\gamma_{t}}{1-\overline{\alpha}_{t}}\boldsymbol{x}_{0}+\frac{\sqrt{\alpha_{t}}\left( 1-\overline{\alpha}_{t-k} \right)}{1-\overline{\alpha}_{t}}\boldsymbol{x}_{t}, \overline{\gamma}=\frac{1-\overline{\alpha}_{t-k}}{1-\overline{\alpha}_{t}}\gamma_{t}
    \label{eq:gamma}
\end{align}
where \( \alpha_{t}=1-\gamma_{t}\) and \( \overline{\alpha}_{t}=\prod_{r=[0,k,..,t]}\alpha_{r}\). 

Eqs. \ref{eq:G-loss}-\ref{eq:D-loss} also require sampling from the network-parameterized denoising distribution $p_{\boldsymbol{\theta}}(\boldsymbol{x}_{t-k}|\boldsymbol{x}_t, \boldsymbol{y})$. A trivial albeit deterministic sample would be the generator output, i.e. $\hat{\boldsymbol{x}}_{t-k} \sim \delta(\boldsymbol{x}_{t-k} - G_{\boldsymbol{\theta}}(\boldsymbol{x}_t, \boldsymbol{y}, t))$. To maintain stochasticity, we instead operationalize the generator distribution as follows: 
\begin{equation} \label{eq:gsample}
    p_{\boldsymbol{\theta}}(\boldsymbol{x}_{t-k}|\boldsymbol{x}_t, \boldsymbol{y}) := q(\boldsymbol{x}_{t-k}|\boldsymbol{x}_t, \tilde{\boldsymbol{x}}_0=G_{\boldsymbol{\theta}}(\boldsymbol{x}_t,\boldsymbol{ y}, t))
\end{equation}
\revhl{where $G_{\boldsymbol{\theta}}$ predicts $\tilde{\boldsymbol{x}}_0$ that is $t/k$ steps away from $\boldsymbol{x}_t$}. Following a total of $T/k$ reverse diffusion steps, the eventual denoised image will be obtained via sampling $\hat{\boldsymbol{x}}_0 \sim p_{\boldsymbol{\theta}}(\boldsymbol{x}_{0}|\boldsymbol{x}_k, \boldsymbol{y}) $.

\subsubsection{Network Architecture} 
\revhl{To synthesize a target-modality image, the reverse diffusion steps parametrized in Eq.\ref{eq:gsample} require guidance from a source-modality image of the same anatomy. However, the training set might include only unpaired images $\boldsymbol{x}_0^A$, $\boldsymbol{x}_0^B$ for the modalities $A$, $B$, respectively. To learn from unpaired training sets, we introduce a cycle-consistent architecture based on non-diffusive and diffusive modules that bilaterally translate between the two modalities.} 

\revhl{\textbf{\emph{Non-diffusive module.}} SynDiff leverages a non-diffusive module to estimate a source image paired with each target image in the training set. A source-image estimate $\tilde{\boldsymbol{y}}^B$ of modality $B$ is produced given a target image $\boldsymbol{x}_0^A$ of modality $A$; and a source-image estimate $\tilde{\boldsymbol{y}}^A$ is produced given a target image $\boldsymbol{x}_0^B$. To do this, two generator-discriminator pairs ($G_{{\boldsymbol{\phi}}^{A}}$,$D_{{\boldsymbol{\phi}}^{A}}$) and ($G_{{\boldsymbol{\phi}}^{B}}$,$D_{{\boldsymbol{\phi}}^{B}}$) with parameters $\boldsymbol{\phi}^{A,B}$ are employed \cite{dar2019image}. The generators produce the estimates $\tilde{\boldsymbol{y}}^{A,B}$ as:
\begin{align}
    \tilde{\boldsymbol{y}}^B = G_{\boldsymbol{\phi}^B}(\boldsymbol{x}_0^A) \notag \\ 
    \tilde{\boldsymbol{y}}^A = G_{\boldsymbol{\phi}^A}(\boldsymbol{x}_0^B) 
\end{align}
A non-saturating adversarial loss is adopted for $G_{{\boldsymbol{\phi}}^{A,B}}$:
\begin{eqnarray} \label{eq:ndG-loss} 
   L_{G_{\boldsymbol{\phi}}} = \mathbb{E}_{p_{\boldsymbol{\phi}}(\boldsymbol{y} | \boldsymbol{x}_0)} [-log(D_{\boldsymbol{\phi}}(\tilde{\boldsymbol{y}}))]
\end{eqnarray}
where $p_{\boldsymbol{\phi}}(y | x_0)$ denotes the network parametrization of the conditional distribution of the source given the target image, and the conditioning input $x_0$ to the discriminator is omitted for brevity. Meanwhile, the discriminators distinguish samples of estimated versus true source images by adopting a non-saturating adversarial loss: \begin{align} \label{eq:ndD-loss} 
   L_{D_{\boldsymbol{\phi}}} = \quad & \mathbb{E}_{q(\boldsymbol{y} | \boldsymbol{x}_0)} [-log(D_{\boldsymbol{\phi}}(\boldsymbol{y}))] \mbox{ } + \notag \\ 
                & \mathbb{E}_{p_{\boldsymbol{\phi}}(\boldsymbol{y} | \boldsymbol{x}_0)} [-log(1-D_{\boldsymbol{\phi}}(\tilde{\boldsymbol{y}}))]
\end{align}
where $q(\boldsymbol{y} | \boldsymbol{x}_0)$ is the true conditional distribution of the source given the target image. Note that, for $D_{{\boldsymbol{\phi}}^{B}}$, $\boldsymbol{y}$ corresponds to $\boldsymbol{x}_0^B$ and the conditioning input is $\boldsymbol{x}_0^A$; whereas for $D_{{\boldsymbol{\phi}}^{A}}$, $\boldsymbol{y}$ corresponds to $\boldsymbol{x}_0^A$ and the conditioning input is $\boldsymbol{x}_0^B$.} 

\revhl{\textbf{\emph{Diffusive module.}} SynDiff then leverages a diffusive module to synthesize target images given source-image estimates from the non-diffusive module as guidance. A synthetic target image $\hat{\boldsymbol{x}}^A$ is produced given $\tilde{\boldsymbol{y}}^{B}$; and a synthetic target image $\hat{\boldsymbol{x}}^B$ is produced given $\tilde{\boldsymbol{y}}^{A}$. To do this, two adversarial diffusion processes are used with respective generator-discriminator pairs ($G_{{\boldsymbol{\theta}}^{A}}$,$D_{{\boldsymbol{\theta}}^{A}}$) and ($G_{{\boldsymbol{\theta}}^{B}}$,$D_{{\boldsymbol{\theta}}^{B}}$) of parameters $\boldsymbol{\theta}^{A,B}$. Starting with Gaussian noise images $\boldsymbol{x}^{A,B}_T$ at time step $T$, target images are synthesized in $T/k$ reverse diffusion steps. In each step, the generators first produce deterministic estimates of denoised target images as noted in Sec. \ref{sec:adp}:  
\begin{align}
\label{eq:getx0}
    \tilde{\boldsymbol{x}}^A_0 = G_{\boldsymbol{\theta}^A}(\boldsymbol{x}^A_t, \boldsymbol{y}=\tilde{\boldsymbol{y}}^B, t) \notag \\
    \tilde{\boldsymbol{x}}^B_0 = G_{\boldsymbol{\theta}^B}(\boldsymbol{x}^B_t, \boldsymbol{y}=\tilde{\boldsymbol{y}}^A, t) 
\end{align}
Afterwards, the denoising distribution for each modality as described in Eq. \ref{eq:gsample} is used to synthesize target images:
\begin{align}
\label{eq:getxhat}
    \hat{\boldsymbol{x}}^A_{t-k} \sim q(\boldsymbol{x}^A_{t-k} | \boldsymbol{x}^A_{t}, \tilde{\boldsymbol{x}}^A_0) \notag \\
    \hat{\boldsymbol{x}}^B_{t-k} \sim q(\boldsymbol{x}^B_{t-k} | \boldsymbol{x}^B_{t}, \tilde{\boldsymbol{x}}^B_0) 
\end{align}
}

\vspace{-0.5cm}
\subsubsection{Learning Procedures}
\revhl{To achieve unsupervised learning, SynDiff leverages a cycle-consistency loss by comparing true target images against their reconstructions. In the diffusive module, reconstructions are taken as synthetic target images $\hat{\boldsymbol{x}}^{A,B}_{0}$. In the non-diffusive module, source-image estimates are projected to the target domain via the generators:
\begin{align}
    \breve{\boldsymbol{x}}_0^A = G_{\boldsymbol{\phi}^A}(\tilde{\boldsymbol{y}}^B) \notag \\
    \breve{\boldsymbol{x}}_0^B = G_{\boldsymbol{\phi}^B}(\tilde{\boldsymbol{y}}^A)
\end{align}
where $\breve{\boldsymbol{x}}_0^{A,B}$ denote the corresponding reconstructions. Afterwards, the cycle-consistency loss is defined as:
\begin{eqnarray} \label{eq:cyclic}
    & L_{\text{cyc}} = \mathbb{E}_{t,q(\boldsymbol{x}_0^{A,B}),q(\boldsymbol{x}_t^{A,B}|\boldsymbol{x}_0^{A,B})} \left[  \lambda_{1\phi} (| \boldsymbol{x}_0^A - \breve{\boldsymbol{x}}_0^A |_1 +  \notag \right. \\ 
    & \left. |\boldsymbol{x}_0^B - \breve{\boldsymbol{x}}_0^B |_1 ) + \lambda_{1\theta}(| \boldsymbol{x}_0^A - \hat{\boldsymbol{x}}_0^A |_1 + | \boldsymbol{x}_0^B - \hat{\boldsymbol{x}}_0^B |_1)  \right]
\end{eqnarray}
where $\lambda_{1\phi,1\theta}$ are the weights for cycle-consistency loss terms from the non-diffusive and diffusive modules respectively, and $\ell_1$-norm of the difference between two images is taken as a consistency measure \cite{dar2019image}. The diffusive and non-diffusive modules are trained jointly without any pretraining procedures. Accordingly, the overall generator loss is:
\begin{align}
\label{eq:ovgen}
    L^{\text{total}}_G = \lambda_{2\phi}(L_{G^A_{\boldsymbol{\phi}}} + L_{G^B_{\boldsymbol{\phi}}}) + \lambda_{2\theta}(L_{G^A_{\boldsymbol{\theta}}} + L_{G^B_{\boldsymbol{\theta}}})  + L_{\text{cyc}}
\end{align}
where $\lambda_{2\phi,2\theta}$ are the weights for adversarial loss terms from the non-diffusive and diffusive modules respectively, and for each modality $L_{G_{\boldsymbol{\phi}}}$ is defined as in Eq. \ref{eq:ndG-loss} and $L_{G_{\boldsymbol{\theta}}}$ is defined as in Eq. \ref{eq:G-loss}. The overall discriminator loss is given as:
\begin{align}
\label{eq:ovdis}
    L^{\text{total}}_D = \lambda_{2\phi}(L_{D^A_{\boldsymbol{\phi}}} + L_{D^B_{\boldsymbol{\phi}}}) + \lambda_{2\theta}(L_{D^A_{\boldsymbol{\theta}}} + L_{D^B_{\boldsymbol{\theta}}}) 
\end{align}
with $L_{D_{\boldsymbol{\phi}}}$ defined as in Eq. \ref{eq:ndD-loss} and $L_{D_{\boldsymbol{\theta}}}$ defined as in Eq. \ref{eq:D-loss}.}

\revhl{During training, the non-diffusive module must be used to produce estimates of source images paired with given target images. During inference, however, the task is to synthesize an unacquired target image given the acquired source image of an anatomy, so only the respective generator within the diffusive module that performs the desired task is needed. For instance, to perform the mapping $A$$\rightarrow$$B$ (i.e., source$\rightarrow$target), $G_{\boldsymbol{\theta}^B}(\boldsymbol{x}_t^B,\boldsymbol{y}^A,t)$ is used where $\boldsymbol{x}_t^B$ is the target-image sample of modality $B$ at time step $t$ and $\boldsymbol{y}^A$ is the acquired source image of modality $A$ provided as input. Inference starts at time step $T$ with a Gaussian noise sample $\boldsymbol{x}_T^B$ drawn from $\mathcal{N}\left( \boldsymbol{0},\boldsymbol{I} \right)$, and the noisy target-image sample produced at the end of each reverse diffusion step is taken as the input target-image sample in the following step. A total of $T/k$ reverse diffusion steps are taken as outlined in Eqs. \ref{eq:getx0}-\ref{eq:getxhat} to attain $\hat{\boldsymbol{x}}_0^B$ at time step $0$ as the synthetic target image.}

\section{Methods}

\subsection{Datasets}  
We demonstrated SynDiff on two multi-contrast brain MRI datasets (IXI\footnote{https://brain-development.org/ixi-dataset/}, BRATS \cite{brats_1}), and a multi-modal pelvic MRI-CT dataset \cite{mr_ct_dataset}. In each dataset, a three-way split was performed to create training, validation and test sets with no subject overlap. \revhl{While all unsupervised medical image translation models were trained on unpaired images, performance assessments necessitate the presence of paired and registered source-target volumes. Thus, in the validation and test sets, separate volumes of a given subject were spatially registered to enable calculation of quantitative metrics. Registrations were implemented in FSL via affine transformation and mutual information loss \cite{fslcitation}. In each subject, each imaging volume was separately normalized to a mean intensity of 1. The maximum voxel intensity across subjects was then normalized to 1 to ensure an intensity range of [0,1]. Cross-sectional images were zero-padded as necessary to attain a consistent 256$\times$256 image size in all datasets prior to modeling.}

\subsubsection{IXI Dataset} \Tone-, \Ttwo-, PD-weighted images from $40$ healthy subjects were analyzed, with ($25$,$5$,$10$) subjects reserved for (training,validation,test). \revhl{\Ttwo~and PD volumes were registered onto \Tone~volumes in validation/test sets.} In each subject, $100$ axial cross-sections with brain tissue were selected. Scan parameters were TE=4.6ms, TR=9.81ms for \Tone; TE=100ms, TR=8178.34ms for \Ttwo; TE=8ms, TR=8178.34ms for PD images; and a common spatial resolution=0.94$\times$0.94$\times$1.2mm$^3$.

\subsubsection{BRATS Dataset} \Tone-, \Ttwo-, Fluid Attenuation Inversion Recovery (FLAIR) weighted brain MR images from $55$ glioma patients were analyzed, with a (training, validation, test) split of ($25$,$10$,$20$) subjects. \revhl{\Ttwo~and FLAIR volumes were registered onto \Tone~volumes in validation/test sets.} In each subject, $100$ axial cross-sections containing brain tissue were selected. Diverse scan protocols were used at multiple institutions. 

\subsubsection{Pelvic MRI-CT Dataset}
Pelvic \revhl{\Tone-, \Ttwo-weighted MRI, and} CT images from $15$ subjects were analyzed, with a (training, validation, test) split of ($9$,$2$,$4$) subjects. \revhl{\Tone~and CT volumes were registered onto \Ttwo~volumes in validation/test sets.} In each subject, $90$ axial cross-sections were selected. \revhl{For \Tone~scans, TE=7.2ms, TR=500-600ms, 0.88$\times$0.88$\times$3mm$^3$ resolution, or TE=4.77ms, TR=7.46ms, 1.10$\times$1.10$\times$2mm$^3$ resolution were prescribed.}
For \Ttwo~scans, TE=97ms, TR=6000-6600ms, 0.88$\times$0.88$\times$2.50mm$^3$ resolution, or TE=91-102ms, TR = 12000-16000ms, 0.88-1.10$\times$0.88-1.10$\times$2.50mm$^3$ resolution were prescribed. For CT scans, 0.10$\times$0.10$\times$3mm$^3$ resolution, Kernel=B30f, or 0.10$\times$0.10$\times$2mm$^3$ resolution, Kernel=FC17 were prescribed. \revhl{To implement synthesis tasks from accelerated MRI scans \cite{kim2018,yurt2021ss}, fully-sampled MRI data were retrospectively undersampled 4-fold in two dimensions to attain low-resolution images at a 16x acceleration rate \cite{kim2018}.} 

 \begin{figure*}[t]
 \centering
 \includegraphics[width=0.965\textwidth]{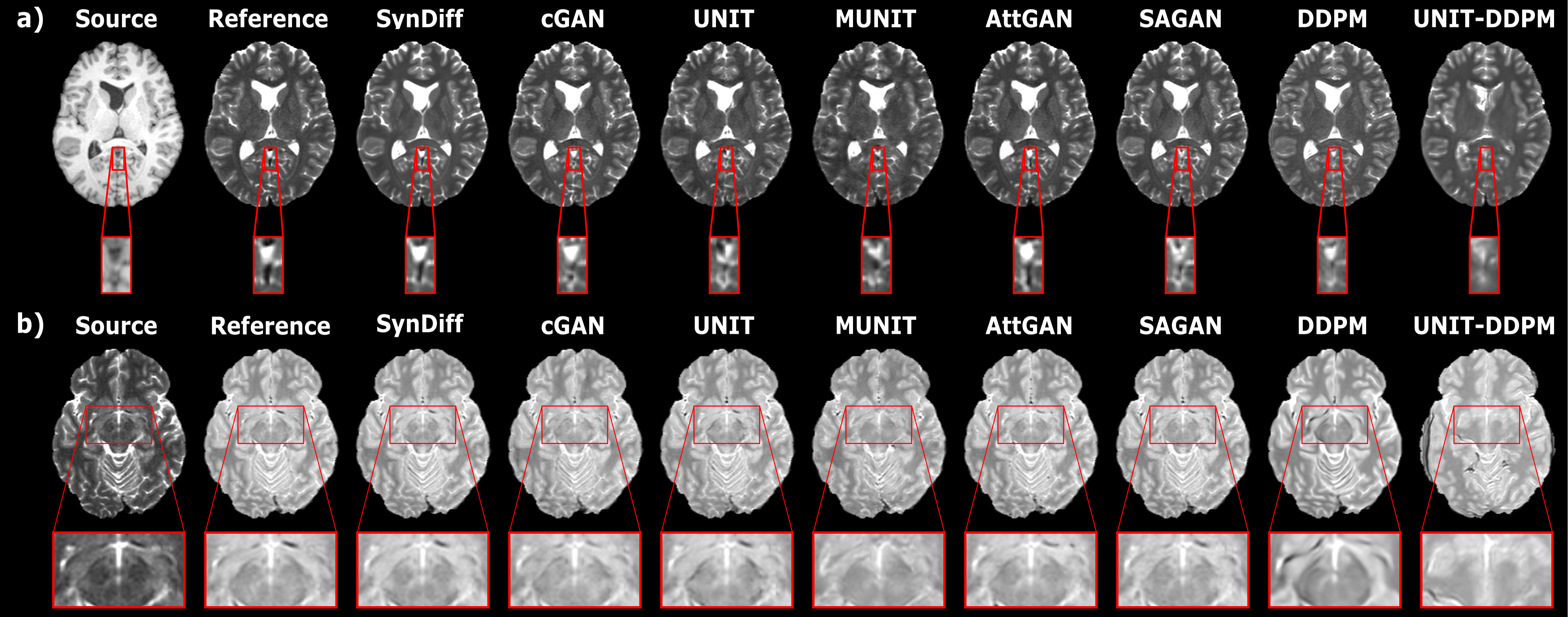}
 \caption{SynDiff was demonstrated on IXI for translation between MRI contrasts. Synthesized images from competing methods are displayed along with the source and the ground-truth target (reference) images for representative \revhl{a) \ToneTtwo,} b) \TtwoPD~tasks. \revhl{Display windows of a) [0 0.65], b) [0 0.80] are used.} Compared to baselines, SynDiff yields lower noise and artifacts, and maintains higher anatomical fidelity.}
 \label{fig:IXI}
 \end{figure*}

\begin{table*}[t]
\caption{Performance for multi-contrast MRI translation tasks in IXI. PSNR (dB) and SSIM (\%) are listed as mean$\pm$std across the test set. Boldface marks the top-performing model in each task.}
\centering
\setlength\tabcolsep{3.5pt}
\resizebox{\textwidth}{!}{%
\begin{tabular}{lcccccccccccc}
\hline
\multirow{2}{*}{}          & \multicolumn{2}{c}{\TtwoTone} & \multicolumn{2}{c}{\ToneTtwo}  & \multicolumn{2}{c}{\PDTone} & \multicolumn{2}{c}{\TonePD} & \multicolumn{2}{c}{\PDTtwo} & \multicolumn{2}{c}{\TtwoPD}  \\ \cline{2-13} 
                           & PSNR      & SSIM     & PSNR      & SSIM & PSNR      & SSIM     & PSNR      & SSIM & PSNR      & SSIM     & PSNR      & SSIM \\ \hline

\multirow{1}{*}{SynDiff} & \textbf{30.42$\pm$1.40} & \textbf{94.77$\pm$1.26} & \textbf{30.32$\pm$1.46} & \textbf{94.28$\pm$1.32} & \textbf{30.09$\pm$1.36} & \textbf{94.99$\pm$1.17}   & \textbf{30.85$\pm$1.56} & \textbf{94.03$\pm$1.12} & \textbf{33.64$\pm$0.86} & \textbf{96.58$\pm$0.36} & \textbf{35.47$\pm$1.15} & \textbf{96.98$\pm$0.36}    \\ \hline
\multirow{1}{*}{cGAN} & 29.22$\pm$1.20 & 93.46$\pm$1.33 & 29.24$\pm$1.26 & 93.01$\pm$1.44 & 28.42$\pm$1.03 & 93.38$\pm$1.19 & 29.92$\pm$1.45 & 93.19$\pm$1.21 & 33.58$\pm$0.75 & 96.46$\pm$0.39 & 34.24$\pm$1.00 & 96.09$\pm$0.47      \\ \hline
\multirow{1}{*}{UNIT}    & 29.17$\pm$1.15 & 93.54$\pm$1.34 & 28.34$\pm$0.98 & 92.02$\pm$1.45 & 28.10$\pm$0.99 & 92.97$\pm$1.21 & 29.29$\pm$1.08 & 92.36$\pm$1.24 & 32.57$\pm$0.65 & 96.22$\pm$0.38 & 34.74$\pm$1.07 & 96.66$\pm$0.39       \\ \hline
\multirow{1}{*}{MUNIT}    & 26.35$\pm$0.88 & 89.78$\pm$1.78 & 26.61$\pm$0.86 & 88.28$\pm$1.90 & 25.99$\pm$0.89 & 89.73$\pm$1.70 & 27.59$\pm$1.02 & 88.71$\pm$1.60  & 29.17$\pm$0.71 & 92.01$\pm$1.01 & 29.80$\pm$0.82 & 91.61$\pm$1.00      \\ \hline
\multirow{1}{*}{AttGAN}& 29.27$\pm$1.38 & 93.74$\pm$1.36 & 28.37$\pm$1.08 & 92.21$\pm$1.45 & 28.02$\pm$1.07 & 92.83$\pm$1.19  & 29.65$\pm$1.42 & 92.98$\pm$1.23 & 32.15$\pm$0.67 & 95.93$\pm$0.44  & 35.11$\pm$1.11 & 96.76$\pm$0.40\\ \hline 
\multirow{1}{*}{SAGAN} & 28.85$\pm$1.26 & 93.38$\pm$1.40 & 29.01$\pm$1.32 & 92.87$\pm$1.43 & 27.93$\pm$1.22 & 93.04$\pm$1.29 & 29.58$\pm$1.51 & 92.76$\pm$1.25 & 32.44$\pm$0.71 & 95.91$\pm$0.46 & 34.75$\pm$0.83 & 96.64$\pm$0.38 \\ \hline
\multirow{1}{*}{DDPM}     &  24.93$\pm$0.69 & 89.49$\pm$1.69 & 28.04$\pm$1.03 & 91.14$\pm$1.58 & 24.95$\pm$0.74 & 89.08$\pm$1.67 & 27.16$\pm$0.95 & 90.45$\pm$1.33 & 30.49$\pm$0.84 & 94.74$\pm$0.69 & 29.67$\pm$0.71 & 93.18$\pm$0.83  \\ \hline
\multirow{1}{*}{UNIT-DDPM}     &  24.01$\pm$0.72 & 86.59$\pm$2.16 & 22.44$\pm$1.26 & 81.64$\pm$3.06 & 23.81$\pm$0.97 &  86.62$\pm$2.44 & 26.81$\pm$1.35 & 88.57$\pm$2.04 & 25.43$\pm$0.49 & 88.08$\pm$1.09 & 25.13$\pm$1.42 & 84.47$\pm$2.53  \\ \hline

\end{tabular}
}
\label{tab:ixi}
\end{table*}

\vspace{-0.3cm}
\subsection{Competing Methods}
We demonstrated SynDiff against several state-of-the-art non-attentional GAN, attentional GAN, and diffusion models. All competing methods performed unsupervised learning on unpaired source and target modalities. For each model, hyperparameter selection was performed to maximize performance on the validation set. \revhl{A common set of parameters that offered near-optimal quantitative performance while maintaining high spatial acuity was selected across translation tasks.} The selected parameters included number of training epochs, learning rate for the optimizer, and loss-term weightings for each model. Additionally, the step size was selected for diffusion models. 

\subsubsection{SynDiff}
\revhl{In the non-diffusive module, generators used a ResNet backbone with three encoding, six residual, and three decoding blocks \cite{he2016deep}; and discriminators used six blocks with two convolutional layers followed by two-fold spatial downsampling. In the diffusive module, generators used a UNet backbone with six encoding and decoding blocks \cite{ronneberger2015u}. Each block had two residual subblocks followed by a convolutional layer. For encoding, the convolutional layer halved feature map resolution and channel dimensionality was doubled every other block. For decoding, the convolutional layer doubled resolution and channel dimensionality was halved every other block. Residual sublocks received a temporal embedding derived by projecting a 32-dimensional sinusoidal position encoding through a two-layer multi-layer perceptron (MLP) \cite{song2020score}. They also received 256-dimensional random latents from a three-layer MLP to modulate feature maps via adaptive normalization \cite{StyleGAN1}. Discriminators used six blocks with two convolutional layers followed by two-fold downsampling, and the temporal embedding was added onto feature maps in each block.} Cross-validated hyperparameters were: 50 epochs, 10$^{-4}$ learning rate, $\mu$=0.5, $T$=1000, a step size of $k$=250, and $T/k$=4 diffusion steps. \revhl{Weights for cycle-consistency and adversarial loss terms were $\lambda_{1\phi,1\theta}$=0.5 and $\lambda_{2\phi,2\theta}$=1, respectively.} Lower and upper bounds on the noise variance schedule were set according to $\overline{\beta}_{\text{min}}$=0.1, $\overline{\beta}_{\text{max}}$=20.

\subsubsection{cGAN}
A cycle-consistent GAN model was considered with architecture and loss functions adopted from \cite{dar2019image}. cGAN comprised two generators with ResNet backbones, and two discriminators with a cascade of convolutional blocks followed by instance normalization. Cross-validated hyperparameters were 100 epochs, 2x10$^{-4}$ learning rate linearly decayed to 0 in the last 50 epochs. Weights for cycle-consistency and adversarial losses were 100 and 1.

 \begin{figure*}[t]
 \centering
 \includegraphics[width=0.965\textwidth]{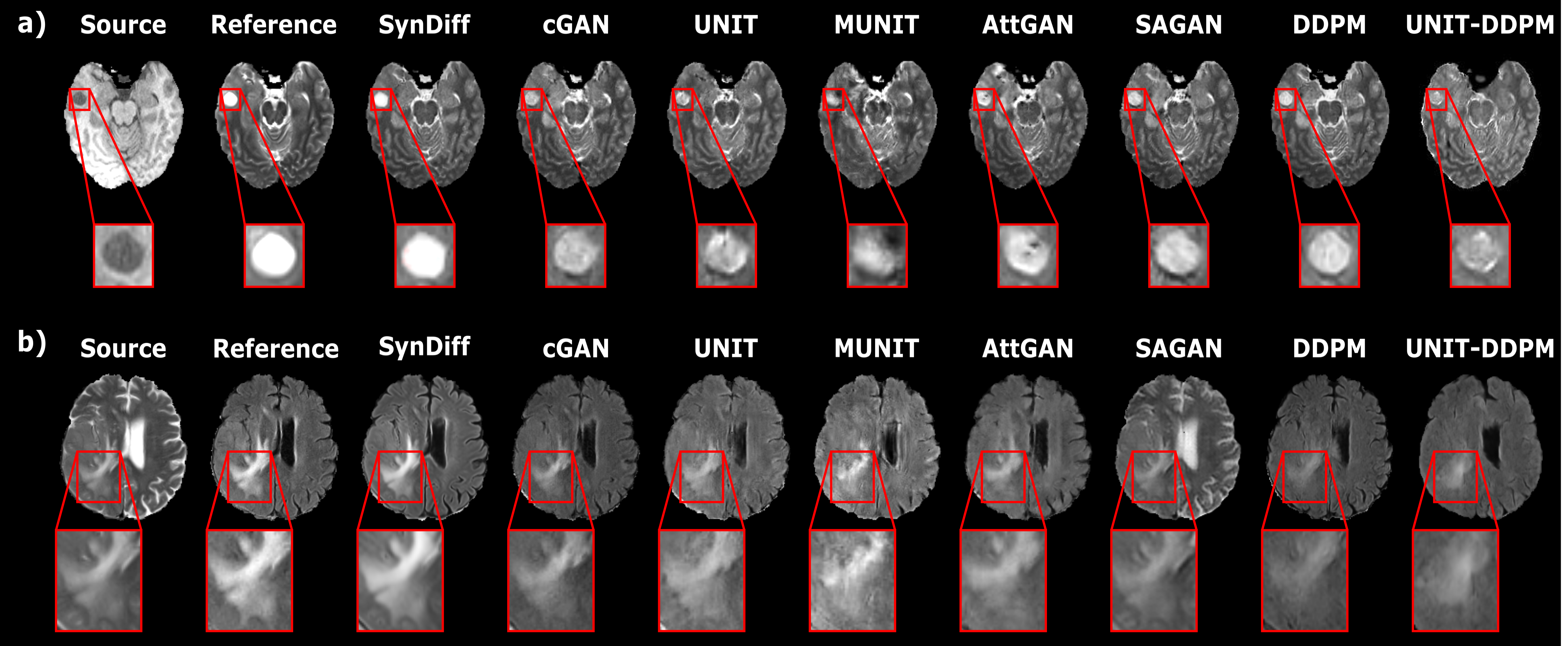}
 \caption{SynDiff was demonstrated on BRATS for translation between MRI contrasts. Synthesized images are displayed along with the source and the ground-truth target (reference) images for representative a) \revhl{\ToneTtwo}, b) \TtwoFlair~tasks. \revhl{Display windows of a) [0 0.75], b) [0 0.80] are used.} SynDiff lowers noise/artifact levels and more accurately depicts detailed structure compared to baselines.}
 \label{fig:BRATS}
 \end{figure*}

\begin{table*}[t]
\caption{Performance for multi-contrast MRI translation tasks in BRATS. PSNR (dB) and SSIM (\%) listed as mean$\pm$std across the test set.} 
\centering
\setlength\tabcolsep{3.5pt}
\resizebox{\textwidth}{!}{%
\begin{tabular}{lcccccccccccc}
\hline
\multirow{2}{*}{}          & \multicolumn{2}{c}{\TtwoTone} & \multicolumn{2}{c}{\ToneTtwo}  & \multicolumn{2}{c}{\FlairTone} & \multicolumn{2}{c}{\ToneFlair} & \multicolumn{2}{c}{\FlairTtwo} & \multicolumn{2}{c}{\TtwoFlair}  \\ \cline{2-13} 
                           & PSNR      & SSIM     & PSNR      & SSIM & PSNR      & SSIM     & PSNR      & SSIM & PSNR      & SSIM     & PSNR      & SSIM \\ \hline

\multirow{1}{*}{SynDiff} & \textbf{28.90$\pm$0.73} & \textbf{93.79$\pm$0.95} & \textbf{27.10$\pm$1.26} & \textbf{92.35$\pm$1.27} & \textbf{26.47$\pm$0.69} & \textbf{89.37$\pm$1.50}   & \textbf{26.45$\pm$1.02} & \textbf{87.79$\pm$1.67} & \textbf{26.75$\pm$1.18} & \textbf{91.69$\pm$1.50} & \textbf{28.17$\pm$0.90} & \textbf{90.44$\pm$1.48}    \\ \hline
\multirow{1}{*}{cGAN}    & 27.41$\pm$0.45 & 92.07$\pm$0.92 & 27.00$\pm$1.11 & 91.90$\pm$1.13 & 26.35$\pm$0.77 & 89.03$\pm$1.51 & 26.44$\pm$0.73 & 85.98$\pm$1.51 & 25.99$\pm$1.30 & 90.02$\pm$1.67 & 27.41$\pm$0.78 & 88.48$\pm$1.45 \\ \hline
\multirow{1}{*}{UNIT}    & 25.76$\pm$0.69 & 87.99$\pm$1.08 & 23.72$\pm$1.15 & 86.62$\pm$1.34 & 26.28$\pm$0.75 & 88.40$\pm$1.46  & 26.41$\pm$0.75 & 86.12$\pm$1.43 & 25.29$\pm$1.34 & 88.41$\pm$1.73 & 26.92$\pm$0.69 & 86.84$\pm$1.43 \\ \hline
\multirow{1}{*}{MUNIT}   & 25.88$\pm$0.73 & 88.16$\pm$1.17 & 23.70$\pm$1.12 & 86.03$\pm$1.34 & 25.08$\pm$0.64 & 86.38$\pm$1.42 & 24.91$\pm$0.76 & 82.73$\pm$1.49 & 24.22$\pm$1.11 & 85.78$\pm$1.39 & 25.26$\pm$0.65 & 83.19$\pm$1.42 \\ \hline
\multirow{1}{*}{AttGAN} & 27.22$\pm$0.47 & 91.87$\pm$0.89 & 26.05$\pm$1.16 & 91.11$\pm$1.36 & 25.59$\pm$0.60 & 87.37$\pm$1.32 & 23.71$\pm$1.13 & 82.12$\pm$2.04 & 24.36$\pm$1.14 & 87.19$\pm$1.52 & 26.56$\pm$0.73 & 86.44$\pm$1.38 \\ \hline 
\multirow{1}{*}{SAGAN} & 26.94$\pm$0.54 & 91.70$\pm$0.96 & 26.60$\pm$1.10 & 91.55$\pm$1.19 & 21.70$\pm$1.02 & 79.82$\pm$3.03 & 23.95$\pm$1.19 & 81.40$\pm$2.44 & 20.33$\pm$1.49 & 79.72$\pm$2.00 & 22.52$\pm$1.02 & 81.02$\pm$1.76  \\ \hline
\multirow{1}{*}{DDPM}    & 27.36$\pm$0.58 & 91.94$\pm$0.96 & 26.34$\pm$1.17 & 91.50$\pm$1.27 & 23.41$\pm$0.64 & 81.55$\pm$2.43 & 24.49$\pm$1.12 & 82.12$\pm$1.97 & 21.23$\pm$1.50 & 82.38$\pm$2.45  & 25.49$\pm$0.60 & 84.71$\pm$1.40 \\ \hline
\multirow{1}{*}{UNIT-DDPM}    & 19.84$\pm$1.54 & 85.92$\pm$2.28 & 23.71$\pm$1.50 & 88.75$\pm$2.49 & 20.31$\pm$0.84 & 79.30$\pm$2.08 & 21.33$\pm$1.18 & 81.80$\pm$1.99 & 20.03$\pm$1.61 & 77.21$\pm$2.03  & 24.15$\pm$1.03 & 82.07$\pm$1.84 \\ \hline
\end{tabular}
}
\label{tab:brats}
\end{table*}

\subsubsection{UNIT}
An unsupervised GAN model that assumes a shared latent space between source-target modalities was considered, with architecture and loss functions adopted from \cite{NIPS2017_dc6a6489}. UNIT comprised two discriminators and two translators with ResNet backbones in a cyclic setup. The translators contained parallel-connected domain image encoders and generators with a shared latent space. The discriminators contained a cascade of downsampling convolutional blocks. Cross-validated hyperparameters were 100 epochs, 10$^{-4}$ learning rate. Weights for cycle-consistency, adversarial, reconstruction losses were 10, 1, and 10.

\subsubsection{MUNIT}
An unsupervised GAN model that assumes a shared content space albeit distinct style distributions for source-target modalities was considered, with architecture and loss functions adopted from \cite{munit}. MUNIT comprised pairs of discriminators, content encoders with ResNet backbones, MLP style encoders, and decoders with ResNet backbones. Cross-validated hyperparameters were 100 epochs, 10$^{-4}$ learning rate. Weights for image, content, style reconstruction, adversarial losses were 10, 1, 1, and 1.

\subsubsection{AttGAN}
A cycle-consistent GAN model with attentional generators \cite{attention_unet} was adopted for unsupervised translation. AttGAN comprised two convolutional attention UNet generators and two patch discriminators \cite{attention_unet}. Cross-validated hyperparameters were 100 epochs, 2x10$^{-4}$ learning rate linearly decayed to 0 in the last 50 epochs. Weights for cycle-consistency and adversarial losses were 100 and 1.

\subsubsection{SAGAN}
A cycle-consistent GAN model with self-attention generators \cite{sagan} was adopted for unsupervised translation. SAGAN comprised two generators based on a ResNet backbone with self-attention layers in the last two residual blocks, and two patch discriminators \cite{sagan}. Cross-validated hyperparameters were 100 epochs, 2x10$^{-4}$ learning rate linearly decayed to 0 in the last 50 epochs. Weights for cycle-consistency and adversarial losses were 100 and 1.

\subsubsection{DDPM}
A recent diffusion model with improved sampling efficiency was considered, with architecture and loss functions adopted from \cite{nichol2021improved}. The source modality was given as a conditioning input to reverse diffusion steps, and cycle-consistent learning was achieved by including non-diffusive modules as in SynDiff. Cross-validated hyperparameters were 50 epochs, 10$^{-4}$ learning rate, $T$=1000, $k$=1, and 1000 diffusion steps. \revhl{A cosine noise schedule was used as in \cite{nichol2021improved}. Weight for cycle-consistency loss was 1}.

\subsubsection{UNIT-DDPM}
\revhl{A recent diffusion model allowing unsupervised training was considered, with architecture and loss functions adopted from \cite{sasaki2021unit}. UNIT-DDPM comprised two parallel diffusion processes for the source and target modalities, where noisy samples from each modality were given as conditioning input to reverse diffusion steps for the other modality. Cross-validated hyperparameters were 50 epochs, 10$^{-4}$ learning rate, $T$=1000, $k$=1, and 1000 diffusion steps. A cosine noise schedule was used \cite{nichol2021improved}. Weight for cycle-consistency loss was 1, and the release time was 1 as in \cite{sasaki2021unit}.}

\begin{figure*}[t]
\centering
\includegraphics[width=0.7\textwidth]{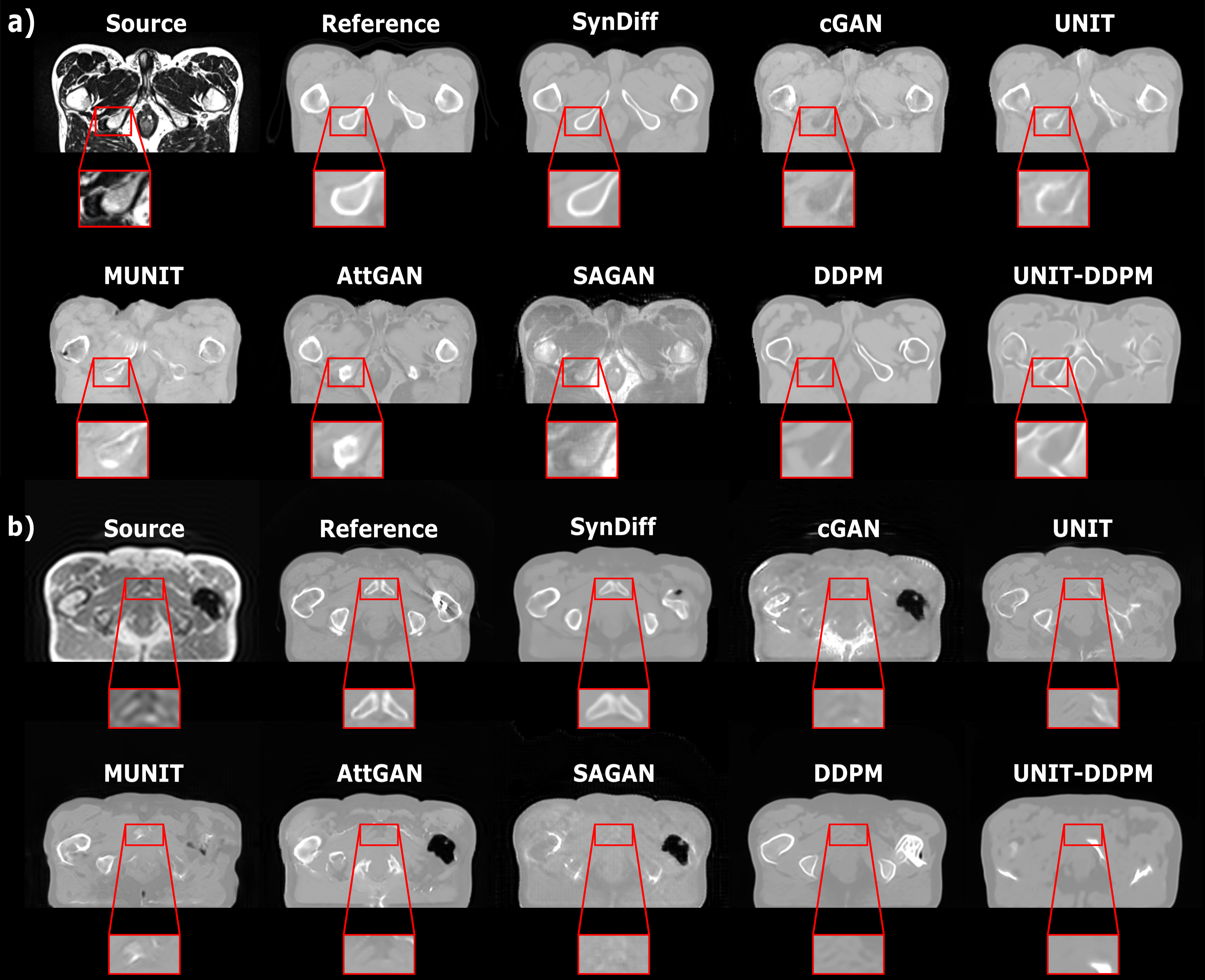}
\caption{SynDiff was demonstrated \revhl{on the pelvic dataset for multi-modal MRI-CT translation.} Synthesized images are displayed along with the source and the ground-truth target (reference) images \revhl{for representative a) \TtwoCT, b) accelerated \ToneCT~tasks. A display window of [0 0.75] is used.} Compared to diffusion and GAN baselines, SynDiff achieves lower artifact levels, and more accurately estimates anatomical structure near diagnostically-relevant regions.}
\label{fig:mr_ct}
\end{figure*}

\vspace{-0.35cm}
\subsection{Modeling Procedures}
All models were implemented in Python using the PyTorch framework. Models were trained using Adam optimizer with $\beta_1$=0.5, $\beta_2$=0.9. Models were executed on a workstation equipped with Nvidia RTX 3090 GPUs. Model performance was evaluated on the test set within each dataset. \revhl{For fair comparison, evaluations of both deterministic and stochastic methods were performed based on a single target image synthesized at each cross section given the respective source image.} Performance was assessed via peak signal-to-noise ratio (PSNR), structural similarity index (SSIM) metrics \revhl{in conditional synthesis tasks where a ground-truth reference is available. For unconditional tasks, Fréchet inception distance (FID) score was utilized to assess the perceptual quality of the generated random synthetic images by comparing their overall distribution to that of actual images.} Prior to assessment, all images were normalized by their mean, and all examined images in a given cross-section were then normalized by the maximum intensity in the reference image. 
Significance of performance differences between competing methods were assessed via non-parametric Wilcoxon signed-rank tests (p$<$0.05).

\begin{table}[t]
\setlength\tabcolsep{4.5pt}
\renewcommand{\arraystretch}{1.3}
\caption{\revhl{Performance for multi-modal MRI-CT translation tasks in the pelvic dataset. PSNR (dB) and SSIM (\%) listed as mean$\pm$std across the test set. `acc.' stands for accelerated.}} 
\centering
\captionsetup{justification = justified,singlelinecheck = false}
\resizebox{1\columnwidth}{!}{
\begin{tabular}{lcccccccc}
\hline
\multirow{2}{*}{}          & \multicolumn{2}{c}{\TtwoCT} & \multicolumn{2}{c}{\ToneCT}  & \multicolumn{2}{c}{acc. \TtwoCT} & \multicolumn{2}{c}{acc. \ToneCT}  \\ \cline{2-9} 
                           & PSNR      & SSIM     & PSNR      & SSIM & PSNR      & SSIM     & PSNR      & SSIM  \\ \hline

\multirow{2}{*}{SynDiff} & \textbf{26.86} & \textbf{87.94} & \textbf{25.16} & \textbf{86.02} & \textbf{26.71} & \textbf{87.32}   & \textbf{25.47} & \textbf{85.00}   \\ 
                         & \textbf{$\pm$0.51} & \textbf{$\pm$2.53} & \textbf{$\pm$1.53} & \textbf{$\pm$2.05} & \textbf{$\pm$0.63} & \textbf{$\pm$2.84}   & \textbf{$\pm$1.09} & \textbf{$\pm$2.10}   \\ \hline
\multirow{2}{*}{cGAN} & 25.07 & 84.91 & 24.11 & 77.81 & 21.24 & 69.62 & 20.35  & 64.73     \\
                      & $\pm$0.17 & $\pm$1.84 & $\pm$1.00 & $\pm$1.84 & $\pm$0.51 & $\pm$0.85 &  $\pm$0.32 & $\pm$1.47     \\ \hline
\multirow{2}{*}{UNIT} & 26.10 & 86.40 & 25.04 & 82.62 & 25.20 & 84.83 &  24.92 &  81.44    \\
 & $\pm$0.49 & $\pm$2.71 & $\pm$0.39 & $\pm$1.52 & $\pm$0.37 & $\pm$1.43 & $\pm$0.39  &  $\pm$1.13   \\ \hline
\multirow{2}{*}{MUNIT} & 22.90 & 77.42 & 24.76 & 79.81 & 23.44 & 77.88 &  24.42 &  79.64    \\
& $\pm$1.05 & $\pm$2.17 & $\pm$0.62 & $\pm$1.20 & $\pm$0.77 & $\pm$2.04 & $\pm$0.34  &  $\pm$1.05    \\ \hline
\multirow{2}{*}{AttGAN} & 23.81 & 74.35 & 24.76 & 82.48 & 23.91 & 76.47 &  21.34 &  67.24   \\ 
 & $\pm$0.18 & $\pm$0.84 & $\pm$1.06 & $\pm$2.49 & $\pm$0.29 & $\pm$0.66 & $\pm$0.51  &  $\pm$1.52   \\ \hline
\multirow{2}{*}{SAGAN} & 21.03 & 67.77 & 23.89 & 77.05 & 19.61 & 61.92 &  23.28 &  70.02    \\ 
 & $\pm$0.33 & $\pm$0.86 & $\pm$1.02 & $\pm$2.87 & $\pm$0.78 & $\pm$0.32 & $\pm$0.96  &  $\pm$2.85    \\ \hline
\multirow{2}{*}{DDPM} & 24.66 & 83.24  & 21.10 & 73.58 & 24.35 & 83.25 & 24.62  & 83.04     \\ 
 & $\pm$0.19 & $\pm$2.62  & $\pm$2.41 & $\pm$7.17 & $\pm$0.47 & $\pm$1.70 & $\pm$0.59  & $\pm$2.40     \\ \hline
\multirow{2}{*}{UNIT-DDPM} & 21.49 & 80.23 & 20.26 & 76.79 & 21.89 & 77.69  & 21.45 & 77.10   \\ 
 & $\pm$0.72 & $\pm$2.69 & $\pm$1.17 & $\pm$1.37 & $\pm$0.77 &  $\pm$3.06 & $\pm$0.23 & $\pm$2.83   \\ \hline
\end{tabular}
}
\label{tab:mr_ct}
\end{table}

\section{Results}
\subsection{Multi-Contrast MRI Translation}
We demonstrated SynDiff for unsupervised MRI contrast translation against state-of-the-art non-attentional GAN (cGAN, UNIT, MUNIT), attentional GAN (AttGAN, SAGAN), and \revhl{regular} diffusion (DDPM, \revhl{UNIT-DDPM}) models. First, experiments were performed on brain images from healthy subjects in IXI. Table \ref{tab:ixi} lists performance metrics for \TtwoTone, \ToneTtwo, \PDTone, \TonePD, \PDTtwo, and \TtwoPD~synthesis tasks. SynDiff yields the highest performance in all tasks (p$<$0.05), except for \PDTtwo~where cGAN performs similarly. On average, SynDiff outperforms non-attentional GANs by 2.2dB PSNR and 2.5$\%$ SSIM, attentional GANs by 1.4dB PSNR and 1.2$\%$ SSIM, and \revhl{regular diffusion models by 5.7dB PSNR and 6.6$\%$ SSIM} (p$<$0.05). Representative images are displayed in Fig. \ref{fig:IXI}. GANs show noise or local inaccuracies in tissue contrast. \revhl{Regular diffusion models suffer} from a degree of spatial warping and blurring. \revhl{UNIT-DDPM shows relatively lower anatomical accuracy, with occasional losses in tissue features.} In comparison, SynDiff yields lower noise and artifacts, and higher accuracy in tissue depiction. 

Next, experiments were conducted on brain images from glioma patients in BRATS. Table \ref{tab:brats} lists performance metrics for \TtwoTone, \ToneTtwo, \FlairTone, \ToneFlair, \FlairTtwo, and \TtwoFlair~tasks. SynDiff again achieves the highest synthesis performance in all tasks (p$<$0.05), except for cGAN that yields similar PSNR in \ToneFlair, and performs similarly in \FlairTone. On average, SynDiff outperforms non-attentional GANs by 1.5dB PSNR and 3.5$\%$ SSIM, attentional GANs by 2.7dB PSNR and 5.0$\%$ SSIM, and \revhl{diffusion models by 4.2dB PSNR and 6.8$\%$ SSIM} (p$<$0.05). Representative images are displayed in Fig. \ref{fig:BRATS}. Non-attentional GANs show elevated noise and artifact levels. Attentional GANs occasionally suffer from leakage of contrast features from the source image (e.g., hallucination of regions with notably brighter or darker signal levels). \revhl{Regular diffusion models show a degree of blurring and feature losses.} Instead, SynDiff generates high-fidelity target images with low noise and artifacts.

\begin{table}[t]
\caption{\revhl{Average training times per cross-section (sec), inference times per cross-section (sec) and memory load (gigabytes)}.} 
\centering
\setlength\tabcolsep{2pt}
\renewcommand{\arraystretch}{1.3}
\captionsetup{justification = justified,singlelinecheck = false}
\resizebox{\columnwidth}{!}{
\begin{tabular}{lcccccccc}
\hline
& SynDiff & cGAN & UNIT & MUNIT & AttGAN & SAGAN & DDPM & UNIT-DDPM\\ \hline
Training & 2.35  & 0.14  &  0.26 & 0.24 & 0.47  & 0.22  & 1.34 & 1.16 \\ \hline
Inference & 0.182  & 0.060  &  0.041 & 0.040 & 0.083  & 0.076  & 85.773 & 52.225 \\ \hline
Memory &  2.12  & 0.77  &  1.31 & 2.26 & 0.86  & 1.08  & 2.95  & 2.77\\ \hline
\end{tabular}
}
\label{tab:times}
\end{table}

\subsection{Multi-Modal Translation}
We also demonstrated SynDiff for unsupervised translation between separate modalities. \revhl{In particular, experiments were performed using SynDiff, non-attentional GAN, attentional GAN, and regular diffusion models on the pelvic dataset for MRI-CT translation. Table \ref{tab:mr_ct} lists performance metrics for \TtwoCT, \ToneCT, accelerated \TtwoCT, and accelerated \ToneCT~synthesis tasks. SynDiff achieves the highest performance in all tasks (p$<$0.05). On average, SynDiff outperforms non-attentional GANs by 2.1dB PSNR and 7.6\% SSIM, attentional GANs by 3.3dB PSNR and 14.4\% SSIM, and diffusion models by 3.6dB PSNR and 7.2$\%$ SSIM (p$<$0.05).} Representative images are displayed in Fig. \ref{fig:mr_ct}. Non-attentional GANs and AttGAN show local contrast losses and artifacts, SAGAN suffers from contrast leakage, and \revhl{regular diffusion models yield} over-smoothing that can cause loss of fine features. \revhl{While UNIT offers higher synthesis performance for some segments near tissue boundaries, particularly around the peripheral body-background boundary, SynDiff has generally higher performance across the image.} Overall, SynDiff synthesizes target images with high anatomical fidelity.

\subsection{Model Complexity}
A practical concern for medical image translation is the computational complexity of the applied models. Table \ref{tab:times} lists the \revhl{training time,} inference time and memory use of competing methods. \revhl{As expected, one-shot GAN models have notably fast training and inference compared to diffusion models. While SynDiff has relatively comparable training times to other diffusion models, its fast diffusion process improves inference efficiency above two-orders-of-magnitude over DDPM and UNIT-DDPM.} In terms of memory utilization, SynDiff has higher demand than cGAN, attentional GANs and UNIT, comparable demand to MUNIT, albeit notably lower demand than DDPM \revhl{and UNIT-DDPM}. Overall, SynDiff offers a more favorable compromise between image fidelity and computational complexity than \revhl{regular diffusion models}.

\subsection{Image Variability}
\revhl{Image translation models involving random noise variables produce stochastic outputs, which can induce variability in target images independently synthesized for a given source image. To assess image variability, we examined target-image samples from competing stochastic methods, SynDiff, MUNIT, DDPM and UNIT-DDPM. For each task, a random selection of 50 cross sections was considered from the test set. For each cross section, 10 target-image samples were synthesized independently given the respective source image. Mean and standard deviation (std.) of performance metrics were computed across 10 samples. On average across cross sections, the std. across samples is less than 0.02dB in PSNR and 0.07\% in SSIM for all methods, except for UNIT-DDPM with std. less than 0.27dB in PSNR and 0.31\% in SSIM. Thus, all stochastic methods have minimal std. values relative to mean values, suggesting limited variability in synthesized target images.}

\begin{figure}[t]
\centering
\includegraphics[width=0.6\columnwidth]{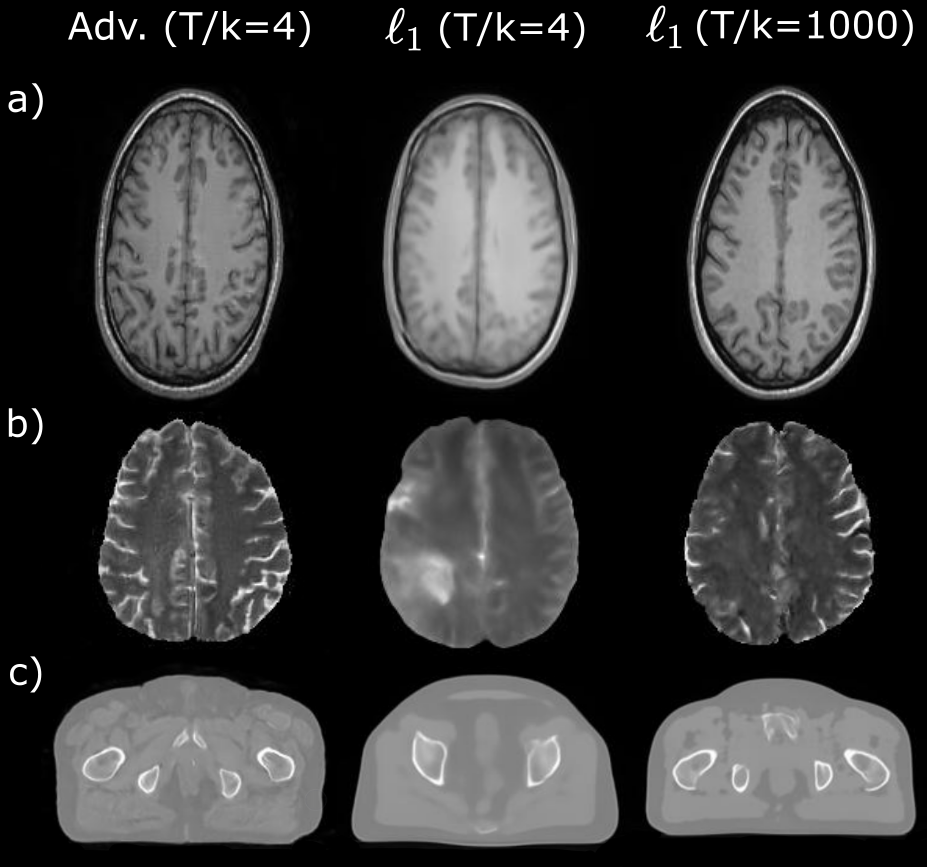}
\caption{The adversarial projector in SynDiff with $T/k$=4 steps was compared against a \revhl{variant model using an $\ell_1$-loss based projector} with $T/k$=4 and $T/k$=1000. Image samples are shown for the unconditional synthesis tasks: a) T\SB{1} in IXI, b) T\SB{2} in BRATS and c) CT in pelvic datasets. \revhl{Display windows of a) [0 0.90], b) [0 0.80], c) [0 0.80] are used.}}
\label{fig:l1proj}
\end{figure}

\begin{table}[t]
\caption{Performance of variant models in unconditional synthesis tasks. FID is listed across the training set.}
\centering
\resizebox{0.650\columnwidth}{!}{%
\begin{tabular}{lccc}
\hline
\multirow{1}{*}{}          & \multicolumn{1}{c}{\Tone (IXI)} & \multicolumn{1}{c}{\Ttwo (BRATS)} & \multicolumn{1}{c}{CT(Pelvic)} \\ \hline
\multirow{1}{*}{Adv. proj. (T/k=4)} & \textbf{30.75} & 75.04 & 58.21 \\ \hline   
\multirow{1}{*}{$\ell_1$ proj. (T/k=4)}  & 141.22 & 96.96 & 107.57     \\ \hline
\multirow{1}{*}{$\ell_1$ proj. (T/k=1000)}  & 52.78  & \textbf{64.66} &  \textbf{54.11}      \\ \hline
\end{tabular}
}
\label{tab:ablation_proj}
\end{table}

\begin{table}[t]
\caption{\revhl{Performance of variant models ablated of adversarial loss, cycle-consistency loss and the diffusive module. PSNR and SSIM listed as mean$\pm$std across the test set.}} 
\centering
\resizebox{0.750\columnwidth}{!}{%
\begin{tabular}{ccccccc}
\hline
\multirow{2}{*}{}          & \multicolumn{2}{c}{\PDTone} & \multicolumn{2}{c}{\ToneTtwo} & \multicolumn{2}{c}{\TtwoCT} \\ \cline{2-7} 
                          & PSNR      & SSIM     & PSNR      & SSIM  & PSNR      & SSIM \\ \hline
\multirow{2}{*}{SynDiff} & \textbf{30.09} & \textbf{94.99}  &  \textbf{27.10} & \textbf{92.35}  &  \textbf{26.86} & \textbf{87.94}      \\
& \textbf{$\pm$1.36}  & \textbf{$\pm$1.17}  & \textbf{$\pm$1.26}  & \textbf{$\pm$1.27} & \textbf{$\pm$0.51}  & \textbf{$\pm$2.53} \\ \hline
\multirow{2}{*}{w/o adv. loss} & 17.69 & 57.97  &  17.87 & 67.87  &  12.48 & 52.36      \\
&  $\pm$0.62 & $\pm$3.45  & $\pm$1.20  & $\pm$2.30 &  $\pm$2.09 & $\pm$3.76 \\ \hline
\multirow{2}{*}{w/o cyc. loss}  &  26.18 &  91.88 & 24.70 & 89.84  &  22.81 & 76.84      \\
&  $\pm$0.73 & $\pm$1.39  & $\pm$1.51  & $\pm$2.21  & $\pm$0.30  & $\pm$2.29 \\ \hline
\multirow{2}{*}{Non-diff. module}  &  28.53 &  93.30 & 26.67 & 90.80  &  22.09 & 80.40      \\
&  $\pm$1.02 & $\pm$1.20  & $\pm$1.05  & $\pm$1.21  & $\pm$1.98  & $\pm$0.32 \\ \hline

\end{tabular}
}
\label{tab:ablation_full}
\end{table}

\begin{table}[t]
\setlength\tabcolsep{2pt}
\renewcommand{\arraystretch}{1.5}
\caption{\revhl{Performance of variant models for varying number of steps T/k and varying loss-term weights ($\lambda_{1\phi},\lambda_{1\theta},\lambda_{2\phi},\lambda_{2\theta}$). PSNR and SSIM listed as mean$\pm$std across the test set.}}
\centering
\resizebox{\columnwidth}{!}{%
\begin{tabular}{lcccccc}
\hline
\multirow{2}{*}{}          & \multicolumn{2}{c}{\PDTone} & \multicolumn{2}{c}{\ToneTtwo} & \multicolumn{2}{c}{\TtwoCT} \\ \cline{2-7} 
                          & PSNR      & SSIM     & PSNR      & SSIM  & PSNR      & SSIM \\ \hline
\vspace{-10pt}
\\ \hline
\multirow{1}{*}{T/k=2}  & 29.47$\pm$1.35 & 94.46$\pm$1.24 & 27.11$\pm$1.31 & 92.48$\pm$1.27    &  26.97$\pm$0.53 & 87.76$\pm$2.56    \\
\multirow{1}{*}{T/k=4} &  30.09$\pm$1.36 & 94.99$\pm$1.17  &  27.10$\pm$1.26 & 92.35$\pm$1.27  &  26.86$\pm$0.51 & 87.94$\pm$2.53      \\
\multirow{1}{*}{T/k=8}  &  29.83$\pm$1.28 & 94.85$\pm$1.18  &  27.24$\pm$1.24 & 92.42$\pm$1.24 &  26.95$\pm$0.59 & 87.95$\pm$2.78      \\ \hline
\vspace{-10pt}
\\ \hline
\multirow{1}{*}{$\lambda_{1\phi}$=0.25}  & 30.04$\pm$1.33 &  94.96$\pm$1.14 & 26.67$\pm$1.20 & 91.89$\pm$1.26    &  26.81$\pm$0.56 & 87.69$\pm$2.66    \\
\multirow{1}{*}{$\lambda_{1\phi}$=0.5}  & 30.09$\pm$1.36 & 94.99$\pm$1.17 & 27.10$\pm$1.26 & 92.35$\pm$1.27    &  26.86$\pm$0.51 & 87.94$\pm$2.53    \\
\multirow{1}{*}{$\lambda_{1\phi}$=1}  & 30.11$\pm$1.29 & 94.97$\pm$1.14 & 27.33$\pm$1.19 & 92.72$\pm$1.17    &  27.07$\pm$0.63 & 88.22$\pm$2.74    \\ \hline

\multirow{1}{*}{$\lambda_{1\theta}$=0.25}  & 29.85$\pm$1.23 & 94.81$\pm$1.14 & 27.18$\pm$1.21 & 92.28$\pm$1.20    &  27.09$\pm$0.58 & 88.01$\pm$2.81    \\
\multirow{1}{*}{$\lambda_{1\theta}$=0.5}  & 30.09$\pm$1.36 & 94.99$\pm$1.17 & 27.10$\pm$1.26 & 92.35$\pm$1.27    &  26.86$\pm$0.51 & 87.94$\pm$2.53    \\
\multirow{1}{*}{$\lambda_{1\theta}$=1}  & 30.06$\pm$1.32 & 94.98$\pm$1.15 & 27.90$\pm$1.25 & 92.03$\pm$1.27    &  27.06$\pm$0.44 & 88.23$\pm$2.52    \\\hline

\multirow{1}{*}{$\lambda_{2\phi}$=0.5}  & 30.12$\pm$1.29 & 95.04$\pm$1.14 & 27.82$\pm$1.20 & 93.12$\pm$1.18   &  26.80$\pm$0.45 & 88.10$\pm$2.46    \\
\multirow{1}{*}{$\lambda_{2\phi}$=1}  & 30.09$\pm$1.36 & 94.99$\pm$1.17 & 27.10$\pm$1.26 & 92.35$\pm$1.27    &  26.86$\pm$0.51 & 87.94$\pm$2.53    \\
\multirow{1}{*}{$\lambda_{2\phi}$=2}  & 29.66$\pm$1.22 & 94.69$\pm$1.15 & 26.99$\pm$1.24 & 91.82$\pm$1.34    &  26.72$\pm$0.53 & 87.69$\pm$2.64    \\ \hline

\multirow{1}{*}{$\lambda_{2\theta}$=0.5}  & 29.97$\pm$1.22 & 94.91$\pm$1.14 & 27.29$\pm$1.24 & 92.59$\pm$1.22   &  26.76$\pm$0.50 & 87.34$\pm$2.55    \\
\multirow{1}{*}{$\lambda_{2\theta}$=1}  & 30.09$\pm$1.36 & 94.99$\pm$1.17 & 27.10$\pm$1.26 & 92.35$\pm$1.27    &  26.86$\pm$0.51 & 87.94$\pm$2.53    \\
\multirow{1}{*}{$\lambda_{2\theta}$=2}  & 29.60$\pm$1.21 & 94.77$\pm$1.14 & 27.20$\pm$1.26 & 92.48$\pm$1.24    &  26.83$\pm$0.48 & 87.97$\pm$2.55    \\\hline
\end{tabular}
}
\label{tab:ablation_param}
\end{table}

\subsection{Ablation Studies}
We conducted a set of ablation studies to systematically evaluate the importance of the main elements in SynDiff. \revhl{To demonstrate the importance of the adversarial diffusion process, we compared the diffusive module in SynDiff based on an adversarial projector against a variant diffusive module based on an $\ell_{1}$-loss based projector for reverse diffusion. The variant module shared the same overall loss function, albeit it ablated the adversarial loss terms for the diffusive generators and discriminators. As such, the remaining loss terms for the diffusive module were based on pixel-wise $\ell_{1}$-loss similar to regular diffusion models. For focused assessment of the diffusive module, demonstrations were performed in unconditional synthesis tasks where guidance from the non-diffusive module was removed from all models.} Synthetic images in representative tasks are displayed in Fig. \ref{fig:l1proj}, and FID scores are listed in Table \ref{tab:ablation_proj}. Compared to the $\ell_1$ projector at $T/k$=4, the adversarial projector at $T/k$=4 substantially improves visual image quality and FID scores over the $\ell_1$ projector at $T/k$=4, while performing competitively with the $\ell_1$ projector at $T/k$=1000. These results demonstrate the utility of adversarial projections for efficient and accurate image sampling during reverse diffusion.    

\revhl{We then examined the contributions of adversarial, cycle-consistent and diffusive learning in SynDiff. A first variant model was constructed by ablating adversarial loss; a second variant model was constructed by ablating cycle-consistency loss; and a third variant model was constructed by ablating the diffusive module to synthesize target images directly using the non-diffusive module. As listed in Table \ref{tab:ablation_full}, SynDiff achieves substantially higher performance than all variants, indicating the importance of each learning strategy. We also assessed the test performance of SynDiff as a function of the number of diffusion steps ($T/k$), and as a function of weights that control the balance between separate loss terms ($\lambda_{1\phi},\lambda_{1\theta},\lambda_{2\phi},\lambda_{2\theta}$). In each case, models were trained across a range of values centered around the parameters selected based on validation performance. As seen in Table \ref{tab:ablation_param}, there are generally minute differences in image quality among variants based on different parameter values. On average across tasks, we find less than 0.2dB PSNR, 0.2\% SSIM difference between the selected and remaining $T/k$ values, and less than 0.3dB PSNR, 0.4\% SSIM difference between the selected and remaining loss-term weights. Overall, these results suggest that SynDiff shows a degree of reliability against parameter variations.} 

\begin{table}[t]
\caption{\revhl{Performance of variant models as mean$\pm$std across the test set. The non-diffusive module was pretrained in variant models. In pretrained-frozen, the non-diffusive module was not updated while training the diffusive module. In pretrained-trained, the non-diffusive module was also updated while training the diffusive module.}}
\centering
\resizebox{0.775\columnwidth}{!}{%
\begin{tabular}{ccccccc}
\hline
\multirow{2}{*}{}          & \multicolumn{2}{c}{\PDTone} & \multicolumn{2}{c}{\ToneTtwo} & \multicolumn{2}{c}{\TtwoCT} \\ \cline{2-7} 
                          & PSNR      & SSIM     & PSNR      & SSIM  & PSNR      & SSIM \\ \hline
\multirow{2}{*}{SynDiff} & \textbf{30.09} & \textbf{94.99}  &  27.10 & 92.35  &  26.86 & 87.94      \\
& \textbf{$\pm$1.36}  & \textbf{$\pm$1.17}  & $\pm$1.26  & $\pm$1.27 & $\pm$0.51  & $\pm$2.53 \\ \hline
\multirow{2}{*}{Pretrained-frozen} & 29.19 & 94.22  &  27.23 & 92.50  &  26.77 & 87.65      \\
&  $\pm$1.28 & $\pm$1.23  & $\pm$1.30  & $\pm$1.31 &  $\pm$0.82 & $\pm$2.81 \\ \hline
\multirow{2}{*}{Pretrained-trained}  &  29.24 &  94.24 & \textbf{27.44} & \textbf{92.75}  &  \textbf{26.97} & \textbf{88.40}      \\
 &  $\pm$1.17 & $\pm$1.21  & \textbf{$\pm$1.23}  & \textbf{$\pm$1.24}  & \textbf{$\pm$0.7}  & \textbf{$\pm$2.74 }\\ \hline
\end{tabular}
}
\label{tab:ablation1_revision}
\end{table}

\begin{table}[t]
\caption{\revhl{Performance of variant models as mean$\pm$std across the test set. In variant models, the non-diffusive module was only trained for $n_{ND}$ epochs while the diffusive module was fully trained. }} 
\centering
\resizebox{0.725\columnwidth}{!}{%
\begin{tabular}{ccccccc}
\hline
\multirow{2}{*}{}          & \multicolumn{2}{c}{\PDTone} & \multicolumn{2}{c}{\ToneTtwo} & \multicolumn{2}{c}{\TtwoCT} \\ \cline{2-7} 
                          & PSNR      & SSIM     & PSNR      & SSIM  & PSNR      & SSIM \\ \hline
\multirow{2}{*}{$n_{ND}=5$}  & 19.88 & 70.31 & 25.26 & 89.99    &  22.59 & 74.59    \\
&$\pm$0.60  & $\pm$2.70  & $\pm$1.08  & $\pm$1.13 & $\pm$0.39  & $\pm$2.44 \\ \hline
\multirow{2}{*}{$n_{ND}=10$} &  27.68 & 92.12  &  26.04 & 91.12  &  24.29 & 81.21      \\
& $\pm$0.70  & $\pm$1.17  & $\pm$1.13  & $\pm$1.26 & $\pm$0.19  & $\pm$2.71 \\ \hline
\multirow{2}{*}{$n_{ND}=25$}  &  29.51 & 94.45  &  26.47 & 91.41 &  26.03 & 86.14      \\
& $\pm$1.14  & $\pm$1.12  &  $\pm$1.33 & $\pm$1.48 & $\pm$0.35  & $\pm$2.52 \\ \hline
\multirow{2}{*}{$n_{ND}=50$}  &  30.09 & 94.99  & 27.10 & 92.35 & 26.86 & 87.94      \\
& $\pm$1.36  & $\pm$1.17  &  $\pm$1.26 & $\pm$1.27 & $\pm$0.51  & $\pm$2.53\\ \hline
\end{tabular}
}
\label{tab:ablation2_revision}
\end{table}

\revhl{Next, we questioned whether SynDiff would benefit from pretraining of the non-diffusive module to improve stability. To address this question, SynDiff was compared against variant models that pretrained the non-diffusive module for 50 epochs to optimize its translation performance, and later combined the pretrained non-diffusive module with a randomly initialized diffusive module. A pretrained-frozen variant trained the combined model while the non-diffusive module was frozen. A pretrained-trained variant trained the combined model while both diffusive and non-diffusive modules were updated. As listed in Table \ref{tab:ablation1_revision}, there are marginal performance changes between SynDiff and variants, with differences less than 0.3dB PSNR and 0.3\% SSIM on average across tasks. These results suggest that the two modules in SynDiff can be jointly trained without notable stability issues or performance losses.}

\revhl{Finally, we assessed the dependence of the diffusive module on the quality of the source-image estimates provided by the non-diffusive module. For this purpose, we trained variant models in which the non-diffusive module was intentionally undertrained to produce suboptimal source-image estimates. Accordingly, the training of the non-diffusive module was stopped early by freezing its weights after a certain number of epochs ($n_{ND}$), while the training of the diffusive module was continued for the full 50 epochs. Table \ref{tab:ablation2_revision} lists performance of variant models across a range of $n_{ND}$ values. Compared to SynDiff at $n_{ND}$=50, we find relatively modest performance differences of 0.7dB PSNR, 1.1\% SSIM at $n_{ND}$=25, and more notable differences of 2.0dB PSNR, 3.6\% SSIM starting at $n_{ND}$=10. These results indicate that while training of the diffusive module shows a degree of reliability against suboptimal source-image estimates, a well-functioning non-diffusive module is key for the performance of the diffusive module in unsupervised medical image translation.}

\section{Discussion}
\revhl{SynDiff adopts adversarial learning for two separate purposes. Adversarial loss in the diffusive module enables accurate reverse diffusion over large step sizes. Meanwhile, adversarial loss in the non-diffusive module enables unsupervised training. In theory, these losses might introduce vulnerability against training instabilities, typically manifested as oscillatory patterns and suboptimal convergence in model performance \cite{mescheder2018training}. To rule out this potential issue, we inspected the validation performance of SynDiff across training epochs. We do not find any notable sign of instabilities as model performance across epochs progresses smoothly towards a convergent point, without abrupt jumps (not reported). We also observe that pretraining of the non-diffusive module does not yield a notable benefit, suggesting that the joint training of diffusive and non-diffusive modules can be performed stably. In cases where instability is suspected during training of SynDiff, stabilization of adversarial components can be achieved via spectral normalization or feature matching \cite{mescheder2018training}.}

\revhl{SynDiff employs a non-diffusive module to provide source-image estimates paired with target images in the training dataset. As such, training of the diffusive module naturally depends on the quality of these source-image estimates. To assess the reliance of the diffusive module on the non-diffusive module, we systematically undertrained the non-diffusive module to produce suboptimal source-image estimates. Note that although the diffusive module was trained with low quality source-image estimates, it was still tested with acquired source images during inference. This creates discrepancy between the distribution of source-image inputs to the diffusive module between the training and test sets. While the diffusive module shows a degree of reliability against moderate discrepancies, its performance degrades as expected under significant discrepancies towards more aggressive levels of undertraining. Thus, a well-functioning non-diffusive module is key for training of the diffusive module.}

\revhl{Generative models for image translation draw samples from the conditional distribution of the target given the source modality. Depending on the use of random variables in this process, generative models can produce either deterministic or stochastic outputs. Among competing methods examined, all GAN models except MUNIT receive only source images to produce deterministic target images. Instead, MUNIT receives random noise variables at intermediate stages. Similarly, all diffusion models initiate sampling on random noise images.} While the influence of the initial sample diminishes across diffusion steps, especially in the presence of a guiding source image, target images are eventually sampled using the reverse transition probabilities. \revhl{Here, we observed that all examined stochastic methods including SynDiff show limited variability across independent target samples synthesized from the same source image, as reflected in performance metrics}. Still, future studies are warranted for an in-depth assessment of the variability of translation estimates and their utility in characterizing uncertainty in diffusion models. 

\revhl{GAN models leverage adversarial learning to achieve high image quality, albeit they can suffer from limited training stability and sample diversity \cite{goodfellow2014generative}. These limitations are particularly burdening in unconstrained image generation tasks, where regular diffusion models have been reported to offer significant benefits \cite{DDPM,nichol2021improved}. Note that training for such unconditional tasks is typically performed on large training sets with highly heterogeneous samples. In contrast, here we consider constrained medical image translation tasks where the model output is anatomically conditioned on a source image, and the size and heterogeneity of the training sets analyzed here are also limited compared to natural image datasets \cite{armanious2019,dar2019image}. As such, benefits of diffusion models in terms of stability and sample diversity might be less discernible. Furthermore, medical images carry higher intrinsic noise than natural images. Regular diffusion models learn a denoising prior using pixel-wise losses, which have lower sensitivity than adversarial losses to fine-grained features including noise \cite{dar2019image}. In turn, DDPM might be prone to limited spatial acuity in synthesizing medical images with significant levels of intrinsic noise \cite{DDM}. This might help explain the less competitive performance of regular diffusion models against GANs in multi-contrast MRI tasks where the target MRI images have higher noise compared to target CT images in multi-modal tasks. Further work is warranted to systematically explore the relative performance of diffusion models against GANs as a function of the size, heterogeneity, and noise levels of medical imaging datasets.}

Starting at random noise, regular diffusion models gradually generate images through repeated denoising over thousands of time steps, resulting in poor sampling efficiency. Here, we proposed to accelerate sampling by leveraging large step sizes coupled with an adversarial projector to capture the complex distribution of reserve transitions. A recent study on MRI reconstruction has reported faster inference by initiating diffusion sampling with an intermediate image from a secondary method \cite{chung2022cvpr}. Improved efficiency in image generation has also been reported for diffusion processes that run in a relatively compact latent space \cite{CardosoLatentDiffusion}. Combining our adversarial projector with these alternative acceleration approaches may result in further speed benefits.

Several technical developments can be further pursued for SynDiff. Here, we demonstrated one-to-one translation tasks between two modalities. When a single source modality does not contain sufficient information to recover the target image, many-to-one translation can be implemented to improve performance \cite{lee2019}. To do this, SynDiff can receive multiple source modalities as conditioning inputs \cite{mmgan,li2019,wang2020}. Second, we considered synthesis tasks in which source and target modalities were unpaired across subjects. When paired source-target images are available, SynDiff can be adapted for supervised training by substituting a pixel-wise in place of cycle-consistency loss and providing actual source images as conditioning input \cite{dar2019image,lyu_arxiv_2022}. Performance improvements might also be viable by expanding the size of training datasets based on a collection of undersampled source- and target-modality acquisitions \cite{yurt2021ss}, or a combination of paired and unpaired source-target modality data \cite{jin2018}. Here the diffusive generators were implemented based on convolutional backbones. Recent studies have reported transformer architectures for improved contextual sensitivity in medical imaging tasks \cite{resvit,TransGAN}. The importance of contextual representations for reverse diffusion remains to be demonstrated, yet an efficient transformer might help enhance the generalization performance to atypical anatomy \cite{slater}. \revhl{Unlike regular diffusion models with slow inference, SynDiff offers a more competitive inference time with GAN models. Yet, its training time is notably higher than GANs, and moderately longer than regular diffusion models due to the computation of added adversarial components and losses. When needed, training efficiency might be improved by parallel execution on multiple GPUs \cite{DiffNvidia}}. The fast conditional diffusion process in SynDiff might also offer performance benefits over GANs in other applications such as denoising and super-resolution \cite{denoisingmri,yang_TMI_2018,transms}.

\revhl{A primary application of SynDiff is imputation of missing scans in multi-contrast MRI and multi-modal imaging. In clinical protocols, a subset of scans are typically omitted due to time constraints, or due to motion artifacts in uncooperative patients \cite{dar2019image}. To maintain the original protocol, omitted scans can then be imputed from acquired scans. Here, high-quality images were synthesized while mapping between native MRI contrasts (e.g., T\SB{1}, T\SB{2}) and mapping MRI to CT. Yet, in other cases, information required to synthesize the target image may not be sufficiently encoded in the source image. For instance, MRI contrasts enhanced with exogenous agents carry distinct information from native contrasts, so it is relatively difficult to synthesize contrast-enhanced MRI images from native MRI contrasts \cite{lee2019}. While CT primarily yields strong contrast for the dense outer bone layers based on X-ray attenuation, MRI shows strong contrast among soft tissues and bone based on tissue magnetization. As such, the primary information on soft tissues needed to synthesize MRI images is scarcely present in CT images, and mapping CT to MRI is a significantly ill-posed problem compared against mapping MRI to CT. Here we observed notably poor performance in CT-to-MRI mapping for all examined methods (not reported). It may be possible to improve translation performance in ill-posed tasks by incorporating multiple source modalities that capture more diverse tissue information \cite{mmgan}.}

\revhl{Another potential application for SynDiff is unsupervised adaptation of learning-based models for downstream tasks such as segmentation and classification across separate domains (e.g., scanners, imaging sites, modalities). When the amount of labeled data is limited in a primary domain, it may be preferred to transfer a model adequately trained in a secondary domain with a large labeled dataset \cite{ae_uda_article,ae_uda_abstract}. However, blind model transfer will incur substantial performance loss given inherent shifts in the data distribution across domains. Assuming that a sufficiently large set of unlabeled images are available in the primary domain, SynDiff can be used to translate these images from primary to secondary domain \cite{uda_diff}. Performance of the transferred model can improve when these translated images are given as input, since their distribution is more closely aligned with secondary-domain images. That said, similar to the case of scan imputation, success in domain adaptation is bounded by the extent of information shared between domains. Downstream models can show suboptimal performance on translated images when information on the secondary domain is not sufficiently encoded in the primary domain.}

\section{Conclusion}
In this study, we introduced a novel adversarial diffusion model for medical image translation between source and target modalities. SynDiff leverages a fast diffusion process to efficiently synthesize target images, and a conditional adversarial projector for accurate reserve diffusion sampling. Unsupervised learning is achieved via a cycle-consistent architecture that embodies coupled diffusion processes between the two modalities. SynDiff achieves superior quality compared to state-of-the-art GAN and diffusion models, and it holds great promise for high-fidelity medical image translation.

\bibliographystyle{IEEETran} 
\bibliography{IEEEabrv,refs}

\end{document}